\documentclass[aps,pre,reprint,10pt,showpacs,showkeys,amsmath,amssymb]{revtex4-1}

\usepackage{amsfonts}
\usepackage{graphicx}
\usepackage{dcolumn}
\usepackage{bm} 
\usepackage{multirow}
\usepackage{topcapt}

\def\gZ{g\mathbf{Z}} \def\gT{g\mathbf{T}} \def\gL{g\mathbf{L}} \def\gP{g\mathbf{P}} \def\gLP{g\mathbf{LP}}
\def\vZ{v\mathbf{Z}} \def\vT{v\mathbf{T}} \def\vL{v\mathbf{L}} \def\vP{v\mathbf{P}} \def\vLP{v\mathbf{LP}}
    \def\LP{\mathbf{LP}}
 
\begin{document}

\title{Backward and covariant Lyapunov vectors and exponents for hard disk systems with a steady heat current}

\author{Daniel P. Truant}
\email[E-mail: ]{d.truant@unsw.edu.au}

\author{Gary P. Morriss}
\email[E-mail: ]{g.morriss@unsw.edu.au}

\affiliation{School of Physics, University of New South Wales, Sydney, New South Wales 2052, Australia}

\date{\today}

\begin{abstract}

The covariant Lyapunov analysis is generalised to systems attached to deterministic thermal reservoirs that create a heat current across the system and perturb it away from equilibrium. The change in the Lyapunov exponents as a function of heat current is described and explained. Both the nonequilibrium backward and covariant hydrodynamic Lyapunov modes are analysed and compared. The movement of the converged angle between the hydrodynamic stable and unstable conjugate manifolds with the free flight time of the dynamics is accurately predicted for any nonequilibrium system simply as a function of their exponent. The nonequilibrium positive and negative $\LP$ mode frequencies are found to be asymmetrical, causing the negative mode to oscillate between the two functional forms of each mode in the positive conjugate mode pair. This in turn leads to the angular distributions between the conjugate modes to oscillate symmetrically about $\pi/2$ at a rate given by the difference between the positive and negative mode frequencies.  

\end{abstract}



\maketitle


\section{Introduction}

   In the chaotic dynamics of the Quasi-One-Dimensional (QOD) system, the Lyapunov exponents indicate the growth or contraction rates of the directions of instability and stability in tangent space \cite{Kane1986, Osel1968, Ruel1979, Ruel1982, EckRue1985, LivPol1986}.
   The stable and unstable directions contain the (covariant) Lyapunov vectors; the stable directions relate to the contraction rates, while the unstable directions relate to the growth rates. 
   More generally, Lyapunov exponents are an important indicator of the degree of chaos in dynamical systems.
   The QOD hard disk system is a subset of more general molecular dynamics (MD) simulations and represents a compromise between the scale of the simulation and the physical limitations on computation. 
     While maintaining the same number of degrees of freedom as a full two-dimensional system, it remains less computationally intensive and allows all relevant features to be analysed \cite{TanMor2003}. 


   The first computationally feasible method of calculating the Lyapunov exponents of chaotic dynamical systems was presented by Benettin et. al. over 30 years ago \cite{BenGal1980,BenGal1976}. 
    The Benettin scheme allowed the calculation of both the Lyapunov exponents and the backward Lyapunov vectors (labelled the BLVs or Gram-Schmidt (GS) vectors \cite{RobMor2008}). 
    The exponents have units of inverse time, thus the largest exponents give information on the fastest microscopic dynamics of the system, while the smallest exponents describe the long-time (or hydrodynamic) behaviour of the system.
    An important feature in the Lyapunov exponent spectrum of some systems is the conjugate paring rule where the shift in the sum of each conjugate pair of exponents remains fixed for all pairs \cite{DetMor1996, ChuTru2010,  Ruel1999, Panj2002}.   
    This illustrates an exact democratic sharing of the total dissipation amongst each two-dimensional hyperbolic sub-space of tangent space formed by conjugate pairs of Lyapunov vectors. 
    

     One of the most interesting features of hard disk systems was the discovery of the Hydrodynamic Lyapunov Modes (HLMs) by Posch \cite{PosHir2000}.
    Lyapunov vectors can show very different features depending on the position of the corresponding exponent in the exponent spectrum. 
   The HLMs are the Lyapunov vectors which form stable delocalised structures across the full breadth of the system and are associated with the smallest magnitude exponents, both positive and negative. 
   While the vector corresponding to the largest-exponent has been analysed since the schemes inception, the existence of HLMs was not immediately recognized, but subsequently a theoretical justification was presented by Eckmann and Gat \cite{EckGat2000}. 
   The HLMs associated with the zero exponents are a separated subspace and comprise the forbidden perturbations in conserved quantities of the system \cite{EckFor2005, RobMor2008, ChuTru2010, MorTru2009}. 
    The vectors associated with the smallest exponents describe the long-time (or hydrodynamic) dynamics of the system, and have been studied by \cite{McNamMar2001-1, WijBei2004, HooPos2002-1, ForHir2004, ForPos2005, EckFor2005, EckGat2000, MilPos2002, TanMor2005}.
    While it is believed that these  HLMs are $k$-vector analogues of the special zero modes \cite{McNamMar2001-1} or Goldstone modes \cite{WijBei2004}, the HLMs can also be analysed as hydrodynamic fields across the perturbations in the tangent space \cite{ChuTru2011}. 
    This interpretation is expanded upon in this paper.


   Although the theoretical formalism of the Lyapunov analysis used covariant (CV) subspaces \cite{EckRue1985, ErsPot1998}, until recently a viable numerical scheme to calculate the covariant Lyapunov vectors was not available. 
   The scheme introduced by Ginelli et. al. \cite{GinPog2007} has allowed the determination of the CV vectors in any system, although other methods have also been proposed \cite{WolSam2007}. 
   The scheme relies upon the GS vectors and the QR decomposition that can be used to obtain them. 
   Although the Ginelli scheme is more computationally intensive, the constrained orthogonality of the Benettin scheme is removed so that the covariant vectors give physically meaningful directions; the stable and unstable manifolds in the tangent space \cite{SzePaz2007, TakGin2009, PazSze2008, BosPos2010, ChaGin2008, YanRad2009, YanRad2010, BosPos2010-1, TruMor2011, PazRod2010,FroHul2012,RomPaz2010,RomPaz2012}. 
   For particle systems where hyperbolicity can be proven or reasonably assumed, the covariant vectors give a numerical tool to investigate the degree of hyperbolicity or to observe its breakdown through the appearance of local tangencies.
   The localization properties of covariant Lyapunov vectors and HLM's for QOD systems at equilibrium have been studied in detail \cite{Morr2012} but it is expected that nonequilibrium particle systems may show a breakdown of hyperbolicity sufficiently far from equilibrium.
   The precursors to this behaviour should be observable using these methods.
   A recent review of the methods and applications of the scheme can be found in the recent special issue \cite{CenGin13}.


   The computer simulation methods for nonequilibrium systems \cite{EvaMor2008} can be grouped into two classes; boundary driven processes and those where an external field drives the system to a steady state. An essential element is a thermostat to remove the heat generated by dissipation. 
   In many cases these two schemes can be shown to be equivalent at least in the linear response region.
   In some cases, such as shear flow, the nonlinear response can be directly connected to rheological problems and thus is physically meaningful.
   For thermal conductivity it is common to use flux boundary conditions \cite{Lep2003} which act as sources of thermalized particles at a fixed temperature but this adds a random element to both the injection of particles and their incoming velocities.
   One advantage of this approach is that all temperature control is done by the external reservoirs.
    In contrast, the deterministic boundary conditions proposed  by Taniguchi and Morriss \cite{TanMor2007a} conserve particle number and allow an energy flux between the system and the reservoir and eliminate the need for a thermostat.
    Here a preliminary analysis of the GS Lyapunov vectors was also reported.
   A comparison of the thermodynamic  results of this model with kinetic theory was made by Kim and Morriss \cite{KimMor09}.
   Later a detailed study of the thermodynamic and scaling properties of the QOD heat conduction system, the boundary effects and a comparison of the phase space contraction and entropy production at equilibrium was reported \cite{MorTru13a}.
   The thermal conductivity is anomolous with an estimated dependence of the heat flux on system size proportional to $N^{1/2}$.
     

   The main focus of the nonequilibrium analysis previously for these systems has been on the thermodynamic properties such as heat conduction and the effects of the contraction of the phase space.
   The Lyapunov analysis (of these systems) has been limited to the changes in the Lyapunov exponent spectrum, and the change in the vectors, for different levels of interaction. 
   There has yet to be a full Lyapunov analysis for nonequilibrium systems, although preliminary attempts have been made using both BLVs \cite{TanMor2007a} and CLVs \cite{BosPos2010-1}. 
   The purpose of this paper is to explore all aspects of the Lyapunov analysis of nonequilibrium QOD hard disk systems; the Lyapunov exponents, vectors both BLV and CLV, and modes as well as their dynamical properties.  
   The system properties we analyse are the temperature profiles, the Lyapunov exponents, the functional form of the Lyapunov vectors (both backward and covariant) and the properties of the modes such as their localisation and angular separation. 
   We look for scaling relations between the imposed system parameters, such as density $\rho$ or heat current $J_Q$, that can be used to systematically organise the properties of these nonequilibrium systems.
   

\section{The nonequilibrium QOD System}\label{sec:qodsystem}

\begin{figure}
       \includegraphics[width=0.45\textwidth]{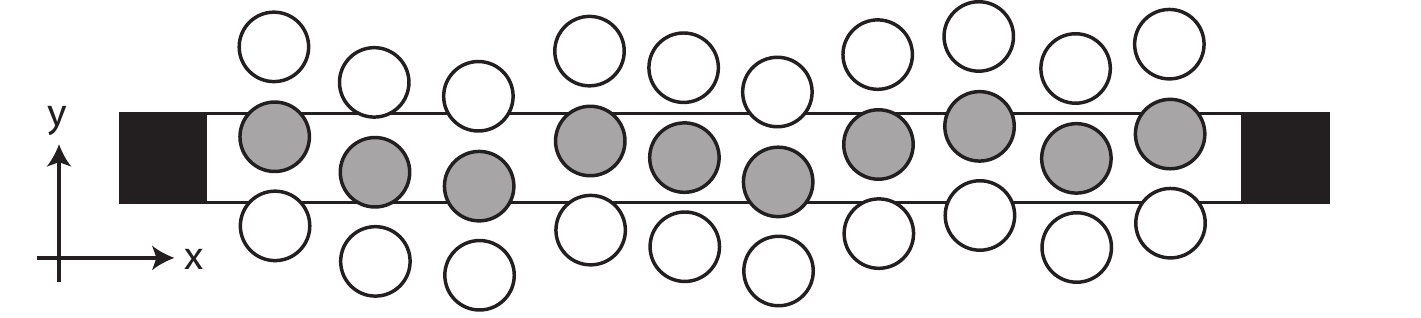}
       \caption{A visual interpretation of the QOD hard disk system, showing here the (H,P) boundary conditions. The shaded disks represent the particles while the unshaded disks represent their periodic images above and below the main channel.}
       \label{fig:QOD}
\end{figure}

   The QOD system encompasses a rectangular space $L_x\times L_y$ populated by $N$ disks, with the condition $L_y < 2\sigma$ ($\sigma$ being the disk diameter, set to unity). This condition ensures the particle ordering remains constant in the $x$ direction. It is usual (although arbitrary) to select $L_y = 1.15\sigma$, thus for a given $N$ a desired density can be found from $\rho = N \sigma^2 / L_xL_y$. These general features are shown in Fig. \ref{fig:QOD}. Crucially, in order to form delocalised structures in the system we need either $L_x$ or $L_y$ to be large. By choosing $L_y = 1.15 \sigma$, we ensure that $L_x \sim N$, and any delocalised structures will develop along the $x$-axis of the system.

   All equilibrium systems analysed were assumed to be isolated and maintained steady time-independent thermodynamic observables. No energy flowed into or out of the system, therefore the thermodynamic state of the system was known exactly. By allowing energy to enter the system from a reservoir at one boundary and leave from the opposite boundary we can consider the effect breaking energy conservation has on the system.

   The QOD we analyse uses (H,P) boundary conditions, giving periodic boundaries in the $y$ direction and hard wall boundaries in the $x$ direction. The system can exchange energy between a reservoir (of momenta $p_{I}$) and a boundary particle (of momenta $p_{xi}$) through a wall collision at the $x=0$ and $x=L_x$ boundaries via 
\begin{equation*}
   p'_{xi} = \epsilon p_I - (1 - \epsilon) p_{xi}
\end{equation*}
   using a nonzero $\epsilon$ ($\epsilon=0.5$ for our analysis). This dynamical coupling ensures that interactions with the thermal reservoirs remain deterministic. 
   Using this interaction a heat current $J_Q$ (an energy current) is maintained through the system, and while no longer in equilibrium a nonequilibrium steady state is achieved and can be analysed.


\section{Equilibrium Results}\label{equilanal}

   A comprehensive analysis of equilibrium hard disk systems has been achieved for both the backward and covariant Lyapunov vectors \cite{Morr2012}. 
   At arbitrary tangent vector in the tangent space at the phase point is given by the difference between the phase trajectory and an infinitesimally perturbed trajectory $\delta \mathbf{\Gamma} = \mathbf{\Gamma}'-\mathbf{\Gamma}$. As the tangent space has the same dimension as the phase space and the tangent vectors span the phase space there are $4N$ tangent vectors. Here the backward Lyapunov vectors and the covariant Lyapunov vectors are analysed.
   
   The $4N$ backward Lyapunov vectors, the $j$th being labeled as $g^{(j)}$, are formed via the Benettin scheme. The $4N$ covariant Lyapunov vectors, the $j$th being labeled as $v^{(j)}$, are formed from the backward Lyapunov vectors via the Ginelli scheme.
   This paper focuses on the BLV and CLV hydrodynamic modes.

\subsection{Backward Lyapunov Modes}

   There are four numerical zero modes in the (H,P) QOD system corresponding to the four zero exponents in the Lyapunov exponent spectrum. The zero mode basis vectors are given by
\begin{eqnarray*}
\gZ^y = \frac{1}{\sqrt{N}} \begin{pmatrix} 0 \\ 1 \\ 0 \\ 0 \end{pmatrix}, \; \gZ^t = \frac{1}{\sqrt{2NT}} \begin{pmatrix} p_x \\ p_y \\ 0 \\ 0 \end{pmatrix} \\ 
\gZ^{p_y} = \frac{1}{\sqrt{N}} \begin{pmatrix} 0 \\0 \\ 0 \\ 1 \end{pmatrix}, \; \gZ^E = \frac{1}{\sqrt{2NT}}  \begin{pmatrix} 0 \\0 \\ p_x \\ p_y \end{pmatrix}.
\end{eqnarray*}   
   Each element is an $N$ dimensional vector; 0 being an $N$ dimensional zero vector, 1 being an $N$ dimensional vector of ones and $p_x$ and $p_y$ are $N$ dimensional vectors where the $i$th entry corresponds to the $i$th particle's $x$ or $y$ momentum. The numerical modes are given as a linear combination of the basis vectors as
\begin{eqnarray*} 
\begin{pmatrix} \gZ^{2} \\ \gZ^{1} \\ \gZ^{-1} \\ \gZ^{-2} \end{pmatrix}^T = \begin{pmatrix} \gZ^{y} \\ \gZ^{t} \\ \gZ^{p_y} \\ \gZ^{E} \end{pmatrix}^T \begin{pmatrix} a & b & 0 & 0 \\ -b & a & 0 & 0 \\ 0 & 0 & 0 & 1 \\ 0 & 0 & 1 & 0 \end{pmatrix}
\end{eqnarray*}   
   The superscripts on the numerical zero modes label the zero modes and are counted outwards from the centre of the spectrum. There are three types of nonzero BLV hydrodynamic Lyapunov modes, transverse (labeled $\gT$), Longitudinal (labeled $\gL$) and momentum proportional (labeled $\gP$). These are given as
\begin{eqnarray*}
\gT^n = \begin{pmatrix} 0 \\ \gamma_n c_n \\ 0 \\ \gamma'_n c_n \end{pmatrix}, \; \gL^n = \begin{pmatrix} \alpha_n s_n \\ 0 \\ \alpha'_n s_n \\ 0 \end{pmatrix}, \; \gP^n = \begin{pmatrix}  \beta_{xn} p_x c_n \\ \beta_{yn} p_y c_n \\  \beta'_{xn} p_x c_n \\  \beta'_{yn} p_y c_n \end{pmatrix}.
\end{eqnarray*}   
   Again each element is an $N$ dimensional vector, the $i$th entry of $c_n = \cos{k_nx_i}$ (similar for $s_n = \sin{k_nx_i}$), and $p_x$ and $p_y$ are the same as defined for the zero modes. The greek symbols represent the (normalised) functional form magnitudes. The $n$ defines the mode number of the mode and also gives the wavevector $k_n = n\pi/L_x$, like the zero modes $n$ increases outwards from the centre of the exponent spectrum, the positive exponent modes give positive $n$ values, the negative exponent modes give negative $n$ values.  The $\gL^n$ and $\gP^n$ modes come in degenerate $\gLP^n$ mode pairs.
\begin{eqnarray*}
\gLP^{n,1} = \sin{\omega_nt} \; \gL^n + \cos{\omega_nt} \; \gP^n \\
\gLP^{n,2} = \cos{\omega_nt} \; \gL^n + \sin{\omega_nt} \; \gP^n
\end{eqnarray*}   
   The $\gT$ modes correspond to the single non degenerate steps in the Lyapunov exponent spectrum, while the $\gLP$ modes correspond to the doubly degenerate exponent steps in the spectrum. For full details of the modes and their properties, the reader is directed to the large literature available for equilibrium backward Lyapunov analysis \cite{MorTru2009,ChuTru2010,MorTru2011}.
       
\subsection{Covariant Lyapunov Modes}   
   
\begin{figure}
  \includegraphics[width=0.45\textwidth]{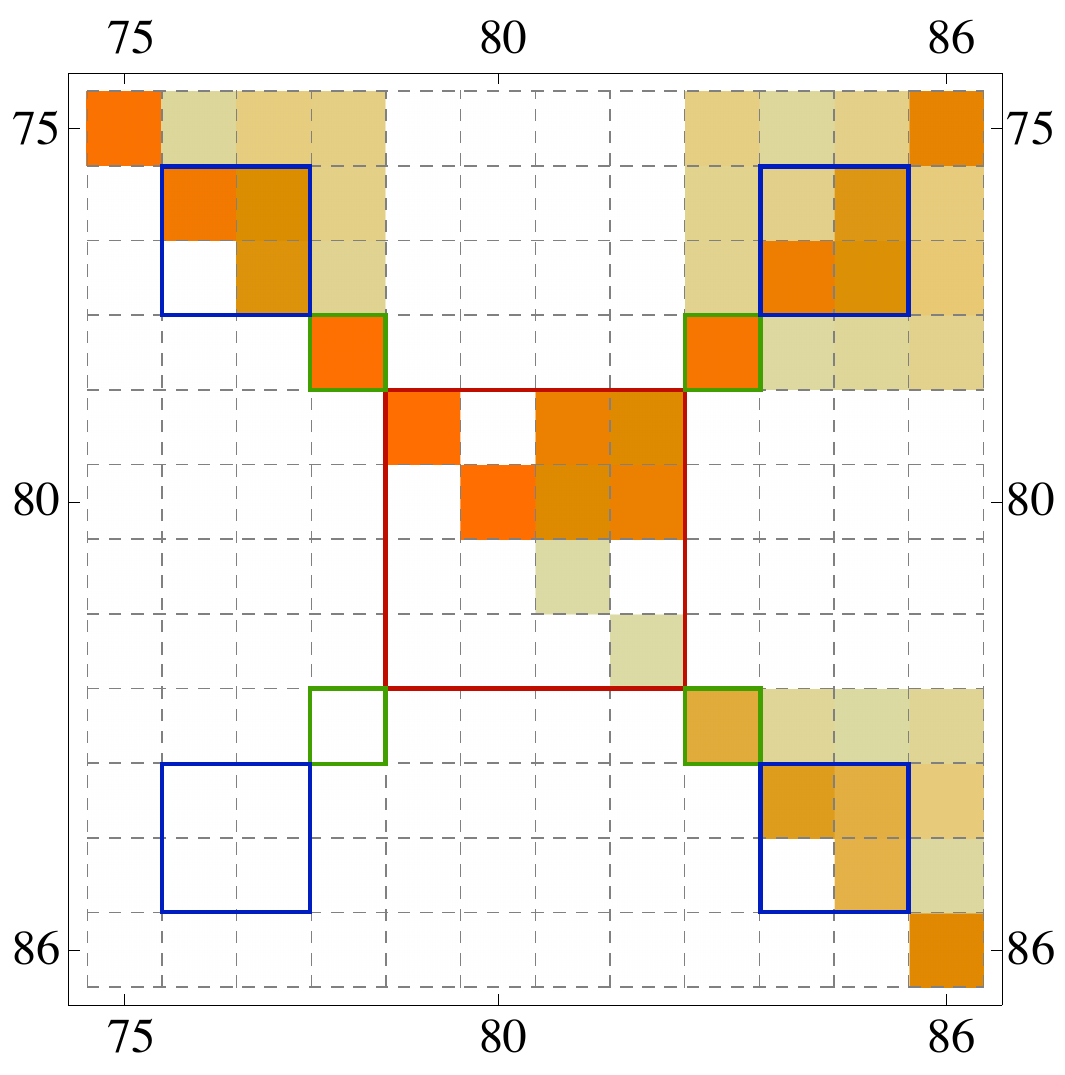}
\caption{The centre of the $C$ matrix, brighter darker entries indicate magnitudes closer to unity. The red square indicates the relevant entries to the $\vZ$ modes, green squares indicate the relevant $\vT$ entries, blue squares the relevant $\vLP$ entries. The labels will be useful in the following analysis.}
\label{fig:Cidentity}
\end{figure}

   The covariant Lyapunov vectors are formed from a linear combination of the backward Lyapunov vectors that precede them. If all $4N$ backward Lyapunov vectors $\{g^{(j)}\}$ are given as column vectors at time step $m$ in a $4N \times 4N$ matrix $G_m$, then the $4N$ covariant Lyapunov vectors $\{v^{(j)}\}$ are given as column vectors in a $4N \times 4N$ matrix $V_m$ as
\begin{eqnarray*}
V_m = G_m C_m.
\end{eqnarray*}   
   $C_m$ is an upper triangular coefficient matrix giving the linear coefficients of the backward Lyapunov vectors that comprise each covariant vector. The covariant Lyapunov modes are given in a similar way to backwards modes, there are four covariant zero modes made of a linear combination of the four backwards zero modes. The centre section of the coefficient matrix at equilibrium is shown in Fig. \ref{fig:Cidentity}. 
   
   As the evolution of the vectors segregates into conjugate mode pairs the coloured squares indicate the relevant $C$ matrix entries for the zero modes, the first transverse conjugate mode pair and the first degenerate $\vLP$ conjugate mode pair. The covariant zero modes are also given via the backward zero mode basis vectors, but are stable and covariant with the dynamics, unlike the backward numerical modes.
\begin{eqnarray*} 
\begin{pmatrix} \vZ^{2} \\ \vZ^{1} \\ \vZ^{-1} \\ \vZ^{-2} \end{pmatrix}^T = \begin{pmatrix} \gZ^{y} \\ \gZ^{t} \\ \gZ^{p_y} \\ \gZ^{E} \end{pmatrix}^T \begin{pmatrix} a & b & 0 & -1 \\ -b & a & -1 & 0 \\ 0 & 0 & 0 & \epsilon \\ 0 & 0 & \epsilon & 0 \end{pmatrix}
\end{eqnarray*}   
   The $4 \times 4$ matrix represents the central $4 \times 4$ block of the $C$ matrix relevant to the zero mode evolution as given by the central red square in Fig. \ref{fig:Cidentity}, which we label as $C_Z$.
   There are three types of nonzero CLV hydrodynamic Lyapunov modes, transverse (labeled $\vT$), Longitudinal (labeled $\vL$) and momentum proportional (labeled $\vP$). Like the backwards modes the $\vL^n$ and $\vP^n$ modes come in degenerate $\vLP^n$ mode pairs.
\begin{eqnarray*}
\vLP^{n,1} = \sin{\omega_nt} \; \vL^n + \cos{\omega_nt} \; \vP^n \\
\vLP^{n,2} = \cos{\omega_nt} \; \vL^n + \sin{\omega_nt} \; \vP^n.
\end{eqnarray*}   
   The first conjugate transverse mode pair ($\vT^1$ and $\vT^{-1}$ is given by the backwards transverse modes as
\begin{eqnarray*}
\begin{pmatrix} \vT^1 \\ \vT^{-1} \end{pmatrix}^T = \begin{pmatrix} \gT^1 \\ \gT^{-1} \end{pmatrix}^T \begin{pmatrix} 1 & \frac{f_T}{\sqrt{f^2_T + 1}} \\ 0 & \frac{1}{\sqrt{f^2_T + 1}} \end{pmatrix}
\end{eqnarray*}   
   where the $2 \times 2$ matrix indicates the relevant $C$ matrix entries as given by the green squares in Fig. \ref{fig:Cidentity} which we label as $C_T$. The $f_T$ function determines the form of the covariant mode evolution. The general $f$ function depends on two sets of variables for each vector, the expansion rates $\{\zeta_i\}$ and the Gram-Schmidt procedural values $\{c_i\}$, and is given by
\begin{eqnarray}\label{equ:ffunc}
 f (m,n,\{\zeta_i\},\{c_i\}) = c_m + \sum^{n}_{i=m+1} \left( c_i \prod_{j=m}^{i-1} \zeta^{-2}_{j} \right)
\end{eqnarray}   
   evaluated between the two backward times steps $m < n$. the subscript $f_T$ indicates the expansion and GS procedural values for the first transverse mode are used. The first degenerate $\vLP$ mode pair is given by
\begin{eqnarray}\label{equ:cforLPs}
\hspace{-1.2cm} \begin{pmatrix} \vLP^{1,2} \\ \vLP^{1,1} \\ \vLP^{-1,1} \\ \vLP^{-1,2} \end{pmatrix}^T = \begin{pmatrix} \gLP^{1,2} \\ \gLP^{1,1} \\ \gLP^{-1,1} \\ \gLP^{-1,2} \end{pmatrix}^T \begin{pmatrix} 1 & 0 & 0 & \frac{f_{LP1}}{\sqrt{f^2_{LP1} + 1}} \\ 0 & 1 & \frac{f_{LP1}}{\sqrt{f^2_{LP1} + 1}} & 0 \\ 0 & 0 & \frac{1}{\sqrt{f^2_{LP1} + 1}} & 0 \\ 0 &  0 &  0 & \frac{1}{\sqrt{f^2_{LP1} + 1}} \end{pmatrix}
\end{eqnarray}   
   here the $4 \times 4$ matrix represents the $C$ matrix entries relevant to the evolution of the first $\vLP$ degenerate mode pair, indicated by the blue squares in Fig. \ref{fig:Cidentity} and labeled as $C_LP$. The subscript on the $f_LP$ function here indicates the expansion and GS procedural values for the first $\LP$ mode are used in the $f$ function of Eq. (\ref{equ:ffunc}).
   
   Unlike the BLVs, which remain an orthogonal set, the covariant vectors evolve naturally to orientate themselves along the stable and unstable manifolds within the tangent space.   The instantaneous angle between any two covariant vectors, $v^{(i)}$ and $v^{(j)}$, is found from the inverse cosine of their inner product
\begin{equation*}
   \theta_{(i,j)} (t) = \cos^{-1}{\left(\big<v^{(i)} \cdot v^{(j)}\big>\right)}
\end{equation*}
   For full details the reader is directed to the equilibrium covariant Lyapunov analysis already undertaken \cite{TruMor2011} 



\section{Temperature Profiles}\label{sec:TP}

\begin{figure*}
                \includegraphics[width=\textwidth]{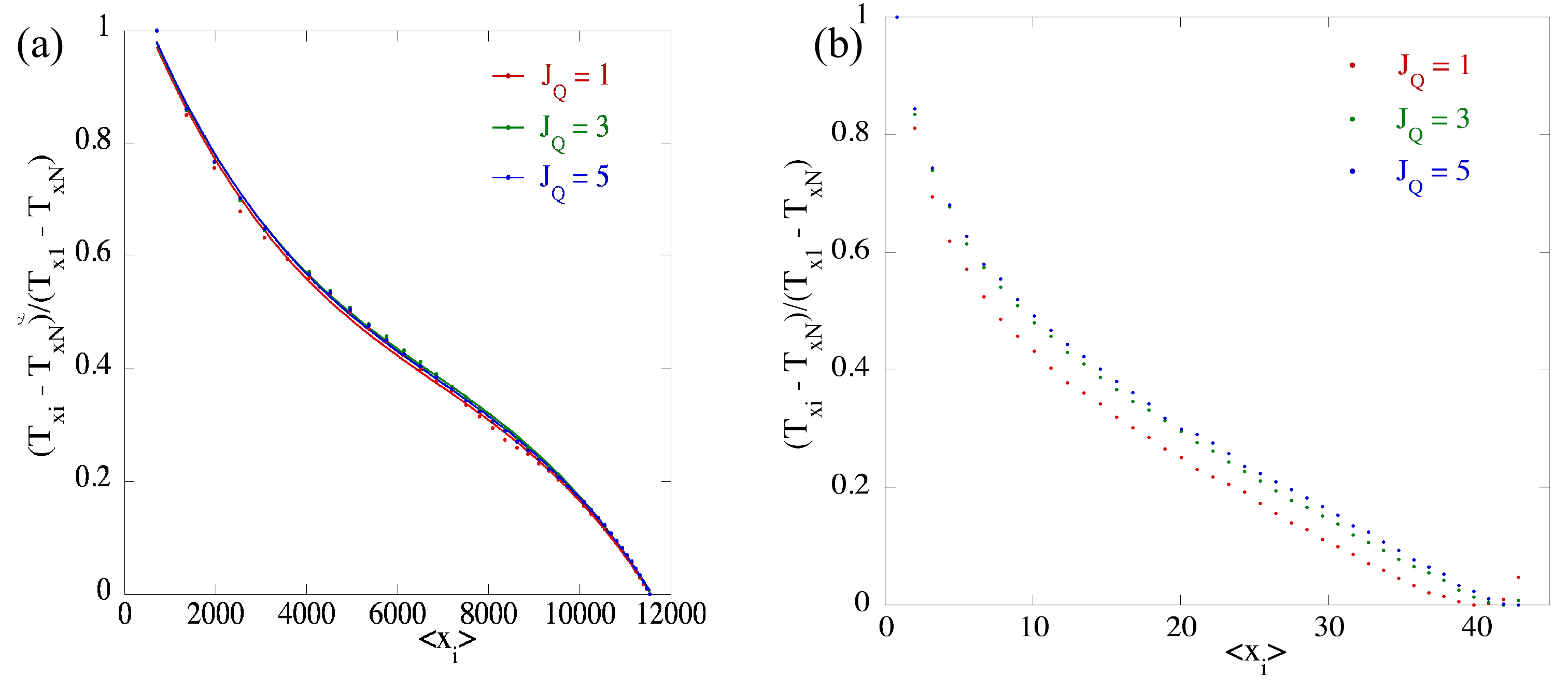}
                \caption{Normalised steady state nonequilibrium temperature profiles in $x$, $(T_{xi}-T_{xN})/(T_{x1}-T_{xN})$, as a function of the average particle position, panel (a) for $\rho = 0.003$, $N=40$ and panel (b) for $\rho = 0.8$, $N=40$. As the temperature profiles are nonlinear, a cubic fitting function is overlaid for $\rho = 0.003$ with fitting parameters given in Table \ref{tab:profilefits}.}
\label{fig:tempprof}
\end{figure*}
   
      The equilibrium and nonequilibrium systems used, all with (H,P) boundary conditions, are shown in Table \ref{tab:heatcurrents}. In all states the initial average temperature of the particles is $\left< T_{i} \right> = 1.0$, but each system has different final steady state temperatures (discussed below). For both high and low density, an equilibrium state (Equ) is compared to a state with wall interaction but no temperature gradient ($J_Q=0$) and states with heat currents of $J_Q = 1$, 3 and 5. An important distinction must be made between the equilibrium system (labeled `Equ') and the $J_Q=0$ system. While both systems have no heat current (from Table \ref{tab:heatcurrents}) the equilibrium system is isolated with $\epsilon =0$, while the $J_Q=0$ system interacts with equal temperature thermal reservoirs with $\epsilon = 0.5$, giving very different system properties as we will see.

\begin{table}[ht]
\caption[Nonequilibrium Thermodynamic System Parameters]{The thermodynamic parameters of the various nonequilibrium systems analysed. For both high density and low density an equilibrium state is compared to; a state with wall interaction but no temperature gradient ($J_Q=0$) and states with heat currents of $J_Q = 1$, 3 and 5. The convergence of the system to a steady state is indicated by balancing the average heat (energy) current through the left and right boundaries, $\Delta E_{L}$ and $\Delta E_{R}$ (the sign defining the direction the current travels at the boundary of the system).}
\renewcommand{\arraystretch}{1.5}
\begin{tabular}{cccccccccc}
\noalign{\hrule height 2pt}
System & $N$ & $\rho$ & $\epsilon$ & $T_{L}$ & $T_{R}$ & $\Delta E_{L}$ & $\Delta E_{R}$ \\
\hline
Equ & 40 & 0.003 & 0.0  & 1.0 & 1.0 & 0.0 & 0.0 \\
$J_Q=0$ & 40 & 0.003 & 0.5  & 1.0 & 1.0 & 0.00001 & -0.00001 \\
$J_Q=1$ & 40 & 0.003 & 0.5  & $\sqrt{500.0}$ & 1.0 & 0.993 & -0.991 \\
$J_Q=3$ & 40 & 0.003 & 0.5  & $\sqrt{1045.0}$ & 1.0 & 3.012 & -3.011 \\
$J_Q=5$ & 40 & 0.003 & 0.5  & $\sqrt{1480.0}$ & 1.0 & 5.045 & -5.045 \\
Equ & 40 & 0.8 & 0.0  & 1.0 & 1.0 & 0.0 & 0.0 \\
$J_Q=0$ & 40 & 0.8 & 0.5  & 1.0 & 1.0 & 0.0001 & -0.0001 \\
$J_Q=1$ & 40 & 0.8 & 0.5  & $\sqrt{3.5}$ & 1.0 & 0.984 & -0.984 \\
$J_Q=3$ & 40 & 0.8 & 0.5  & $\sqrt{7.0}$ & 1.0 & 3.017 & -3.015 \\
$J_Q=5$ & 40 & 0.8 & 0.5  & $\sqrt{9.6}$ & 1.0 & 4.952 & -4.952 \\
\noalign{\hrule height 2pt}\end{tabular}
\label{tab:heatcurrents}
\end{table}
   
   Previous work on the nonequilibrium Lyapunov analysis of QOD systems \cite{TanMor2007a} focused on the alteration of the Lyapunov exponent spectrum with the application of a heat current, finding a nonzero negative shift in the negative $\gLP$ mode exponents compared to the positive $\gLP$ mode exponents, as well as a change in the average functional form of the $\gZ^{-2}$ zero mode (associated with energy conservation, which is broken away from equilibrium). The form of the $\gLP$ modes was also presented with their period of oscillation related to the momentum autocorrelation function for different values of the interaction parameter $\epsilon$.
   
   The temperature profiles of isolated equilibrium systems ($\epsilon = 0$) and nonequilibrium systems with wall-interacting boundary conditions ($\epsilon = 0.5$) have been compared \cite{MorTru2012}. Like the equilibrium and $J_Q=0$ systems analysed here the equilibrium systems were isolated and therefore had no heat current by definition, while the nonequilibrium wall interacting systems had $T_L=T_R=1$, also giving $J_Q=0$ (after convergence). The interaction with the wall was seen to cause a large nonlinear effect towards the boundaries in the temperature profile of the system. A relation was found between an energy balance within the system and the convergence to a steady state, although out of equilibrium. Systems out of equilibrium were seen to produce entropy near the system-reservoir boundary, which flowed into the reservoirs. This flow was negative for the same first $\sim$10 particles regardless of system size, while the entropy production was positive for all particles along the system. Only one nonequilibrium system analysed here at each density has a similar configuration of $J_Q=0$ as used in \cite{MorTru2012} (see Table \ref{fig:tempprof}).
      
   Fig. \ref{fig:tempprof} (a) shows the normalised $x$ temperature profiles  $(T_{xi}-T_{xN})/(T_{x1}-T_{xN})$ of the low density systems ($\rho = 0.003$) analysed as a function of average particle position for heat currents $J_Q = 1$, 3 and 5. The form of the temperature profiles is seen to be independent of the strength of the heat current, only the magnitude increases (which is removed in Fig. \ref{fig:tempprof} (a)). The temperature profiles can be described almost exactly by a cubic function, not a linear function, in average particle position. Fitting parameters of the cubic function for each system lie remarkably close, over 12 orders of magnitude, and are given in Table \ref{tab:profilefits}.

\begin{table}[ht]
\caption[Temperature Profile Fit Parameters]{The fitting parameters of the function $T_x(\left<x_i\right>) = c_0 + c_1 \left<x_i\right> + c_2 \left<x_i\right>^2 + c_3 \left<x_i\right>^3$ for the low density ($\rho=0.003$) temperature profiles of Fig. \ref{fig:tempprof} (a). Over 12 orders of magnitude the parameters are remarkably similar for each system.} 
\renewcommand{\arraystretch}{1.5}
\begin{tabular}{ccccccc}
\noalign{\hrule height 2pt}
Parameter & $J_Q=1$ & $J_Q=3$ & $J_Q=5$ \\
\hline
$c_0$ & 1.1061 & 1.1119  & 1.1155 \\
$c_1$ & -2.0793  $\times 10^{-4}$ & -2.0722  $\times 10^{-4}$  & -2.0813  $\times 10^{-4}$ \\
$c_2$ & 2.2142 $\times 10^{-8}$ &  2.2375 $\times 10^{-8}$ & 2.2146 $\times 10^{-8}$ \\
$c_3$ & -1.0754 $\times 10^{-12}$ &  -1.1038 $\times 10^{-12}$ & -1.0781 $\times 10^{-12}$ \\
$R$ & 0.99927 & 0.99963  & 0.99969 \\
\noalign{\hrule height 2pt}\end{tabular}
\label{tab:profilefits}
\end{table}

\begin{figure*}
                \includegraphics[width=\textwidth]{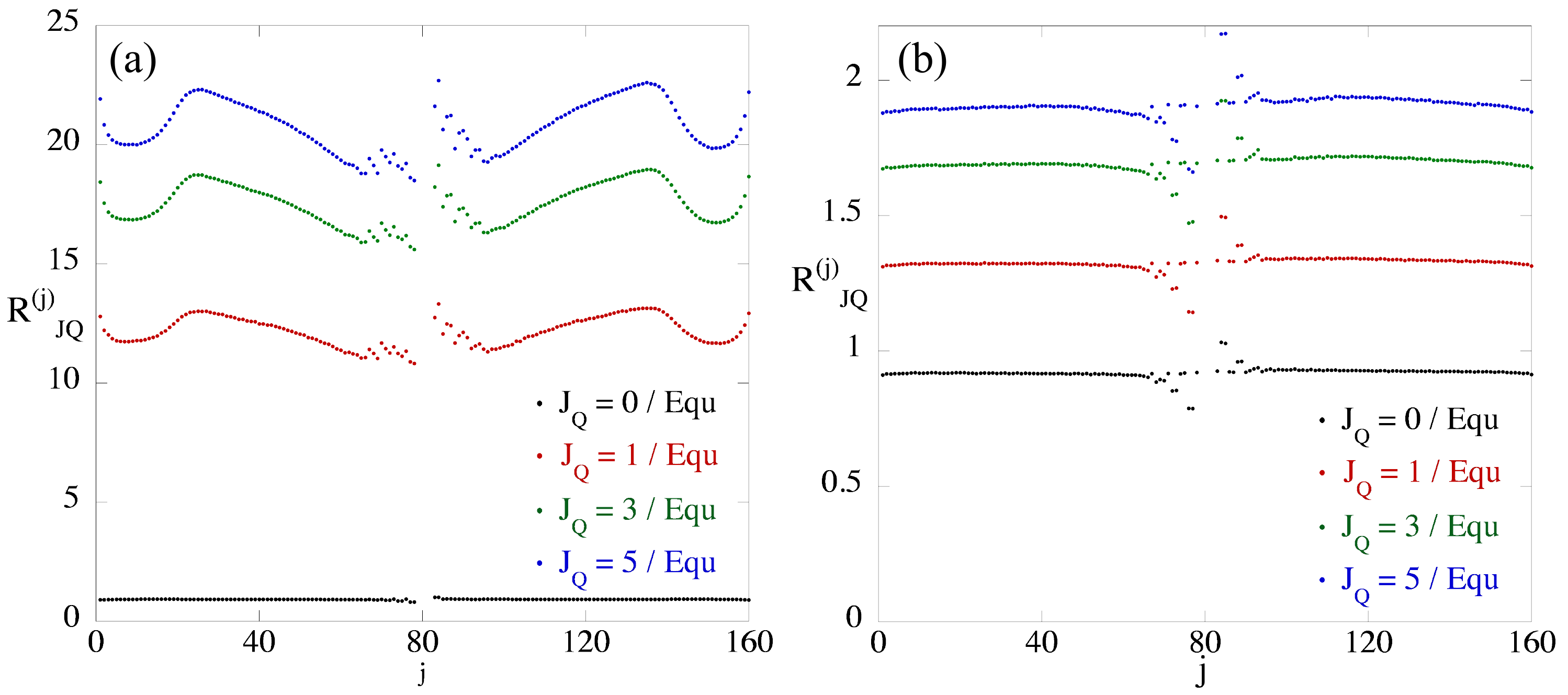}
        \caption{Pointwise ratio of the Lyapunov exponents for each of the nonequilibrium systems compared to the equilibrium system $ R^{(j)}_{J_Q} = \lambda^{(j)}_{J_Q} / \lambda^{(j)}_{Eq}$, panel (a) for $\rho = 0.003$, $N=40$ and panel (b) for $\rho = 0.8$, $N=40$. Almost uniform pointwise scaling with heat current $R^{(j)}_{J_Q} = R_{J_Q}$ is seen for $\rho=0.8$.}
\label{fig:expratios}
\end{figure*}

   Fig. \ref{fig:tempprof} (b) shows that for high density systems analysed here, this functional form no longer holds; the normalised temperature profiles depend strongly on the magnitude of the heat current. As the heat current increases, the distortion of the temperature profile via the introduction of the wall interaction (as seen in \cite{MorTru2012}) is reduced and the profile becomes closer to the cubic kinetic region form.

   One result found for the nonequilibrium $J_Q=0$ systems was the absence of a thermodynamic limit for the temperature profile. An increase in system size leads to a decrease in the temperature in the middle of the system. 
   Despite this, it has been shown the thermodynamic limit for the tangent space dynamics is easily met before the system size reaches $N=50$ \cite{MorTru2011}, therefore we can say with some confidence that the tangent space dynamics obtained is representative regardless of system size.
   Away from equilibrium the density profile associated with the temperature profile is limited by the maximum local density possible before a phase transition. This limits the maximum system size that can be achieved for any nonzero temperature gradient.

\section{Non-Equilibrium Lyapunov Exponent Spectrum}\label{sec:NELES}

   Changing from $\epsilon=0$ to $\epsilon=0.5$, and subsequently introducing a heat current, leads to an increase in the values of the exponents.
   To see exactly how this affects the spectrum of exponents Fig. \ref{fig:expratios} compares the equilibrium and nonequilibrium exponents pointwise across the spectrum, $\lambda^{(j)}_{J_Q} / \lambda^{(j)}_{Eq}$ (finding their direct difference $\lambda^{(j)}_{J_Q} - \lambda^{(j)}_{Eq}$ does not indicate anything useful).
      
   For the high density state, Fig. \ref{fig:expratios} (b), the exponent spectrum scaling ratio,  $ R^{(j)}_{J_Q} = \lambda^{(j)}_{J_Q} / \lambda^{(j)}_{Eq}$, is almost exactly uniform pointwise across the spectrum $R^{(j)}_{J_Q} = R_{J_Q}$ (accept for a few distinct points in the central mode region which we discuss below). This scaling ratio indicates a power law relationship of the nonequilibrium exponents as a function of the heat current, given to high accuracy as
\begin{equation}\label{equ:rjratio}
   R_{J_Q} \cong 0.91 + 0.42 (J_Q)^{1/2}.
\end{equation}
   The low density spectrums, Fig. \ref{fig:expratios} (a), do not scale as uniformly as the high density spectrums. A definite nonuniform shape to the scaling ratio can be seen along the spectrum (with the shape consistent for all heat currents). Not distinguishable in Fig.  \ref{fig:expratios} (a), the shape of the scaling ratio of the $J_Q=0$ system is the inverse of the nonzero heat current ratios. Despite this, the low density scaling ratios do follow a similar power law in the heat current $J_Q$ as the high density states.
 
\begin{figure*}
                \includegraphics[width=\textwidth]{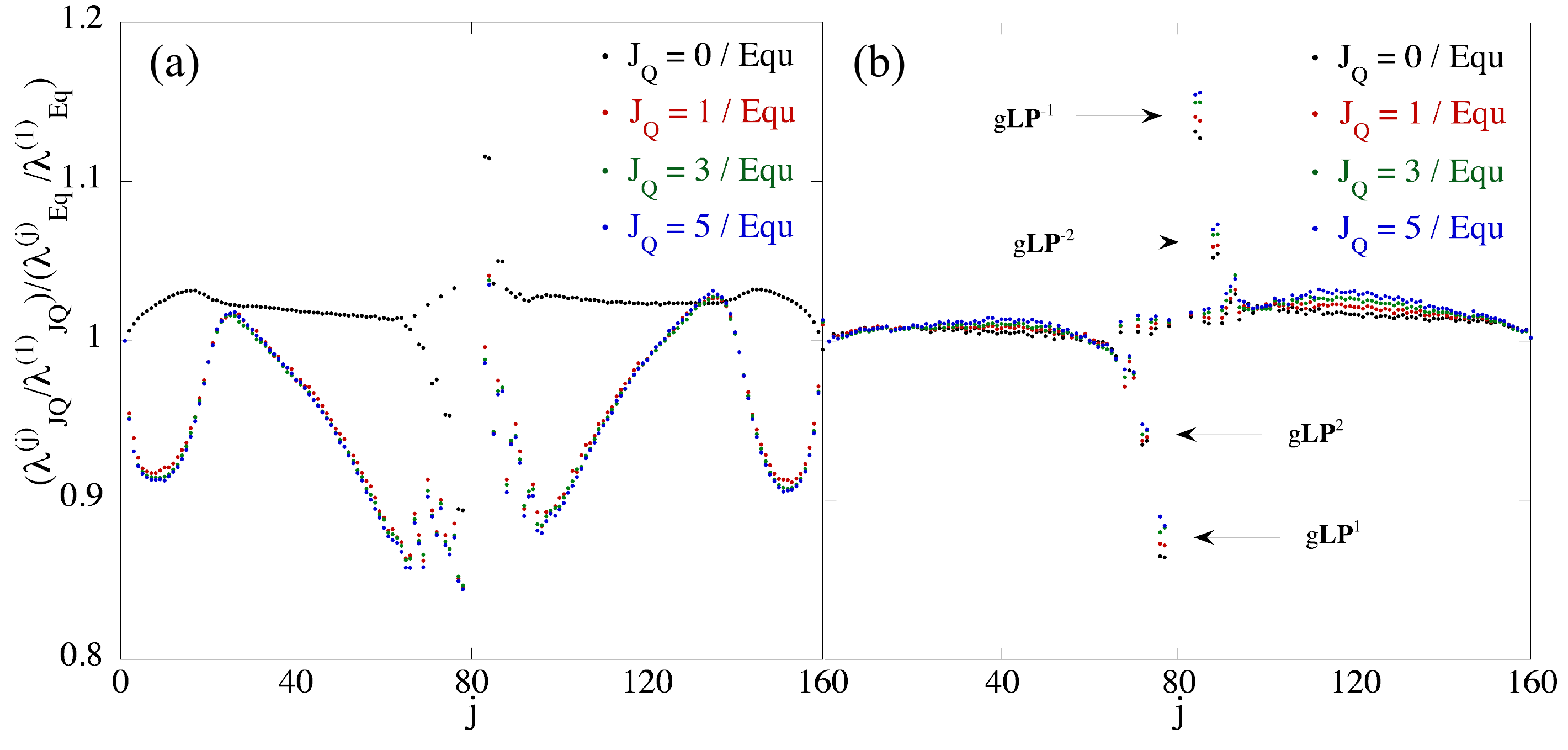}
\caption{Normalised pointwise ratios between the equilibrium and nonequilibrium exponent spectrums $(\lambda^{(j)}_{J_Q}/\lambda^{(1)}_{J_Q}) / (\lambda^{(j)}_{Eq}/\lambda^{(1)}_{Eq})$,  panel (a) for $\rho = 0.003$, $N=40$ and panel (b) for $\rho = 0.8$, $N=40$. Each full spectrum has been normalised with respect to their largest exponent $\lambda^{(j)} \rightarrow \lambda^{(j)}/\lambda^{(1)}$.}
\label{fig:normexpsratio}
\end{figure*}

   Each exponent spectrum can be normalised with respect to their largest exponent $\lambda^{(j)} \rightarrow \lambda^{(j)}/\lambda^{(1)}$. This allows us to see the shape of the exponent spectrum independent of the magnitude which increases due to the changes in heat current. In order to see any structures in the spectrums which remain after removing the scaling due to the heat current, Fig. \ref{fig:normexpsratio} shows each \emph{normalised nonequilibrium} exponent divided by the \emph{normalised equilibrium} exponent $(\lambda^{(j)}_{J_Q}/\lambda^{(1)}_{J_Q}) / (\lambda^{(j)}_{Eq}/\lambda^{(1)}_{Eq})$ pointwise across the spectrum. If the spectrums were identical save for a constant ratio (which is seen for the high density exponents of Fig. \ref{fig:expratios} (b)), we expect a constant unity value pointwise across all $4N$ components $(\lambda^{(j)}_{J_Q}/\lambda^{(j)}_{Eq})/R^{(1)}_{J_Q} = R_{J_Q}/R_{J_Q} = 1$.

   This behaviour is almost exactly what is seen for the high density case in Fig. \ref{fig:normexpsratio} (b). The majority of the normalised exponent ratios are roughly unity, while there is a small constant increase for the exponents in the negative continuous region (the right half of the figure) and there are a series of clear non-unity ratios (indicted in Fig. \ref{fig:normexpsratio} (b) for the modes they correspond to, which will be discussed below). 
   
   In comparison to the high density states there is a clear non-uniformity of the normalised equilibrium and nonequilibrium ratios for the low density case (Fig. \ref{fig:normexpsratio} (a)), independent of heat current and reversed for the $J_Q=0$ system. The large perturbation effect seen for the largest magnitude exponents can be justified. For low density systems the free flight mapping can dominate the tangent space dynamics (via the large $\tau_n$), but the reorthogonalisation is only performed during collision mappings, therefore the largest vectors can grow in directions away from their `correct' directions during the free flight. This can artificially lower the numerical vectors growth, possibly lowering their exponent. 
   
\begin{figure}[ht]
   \includegraphics[width=0.45\textwidth]{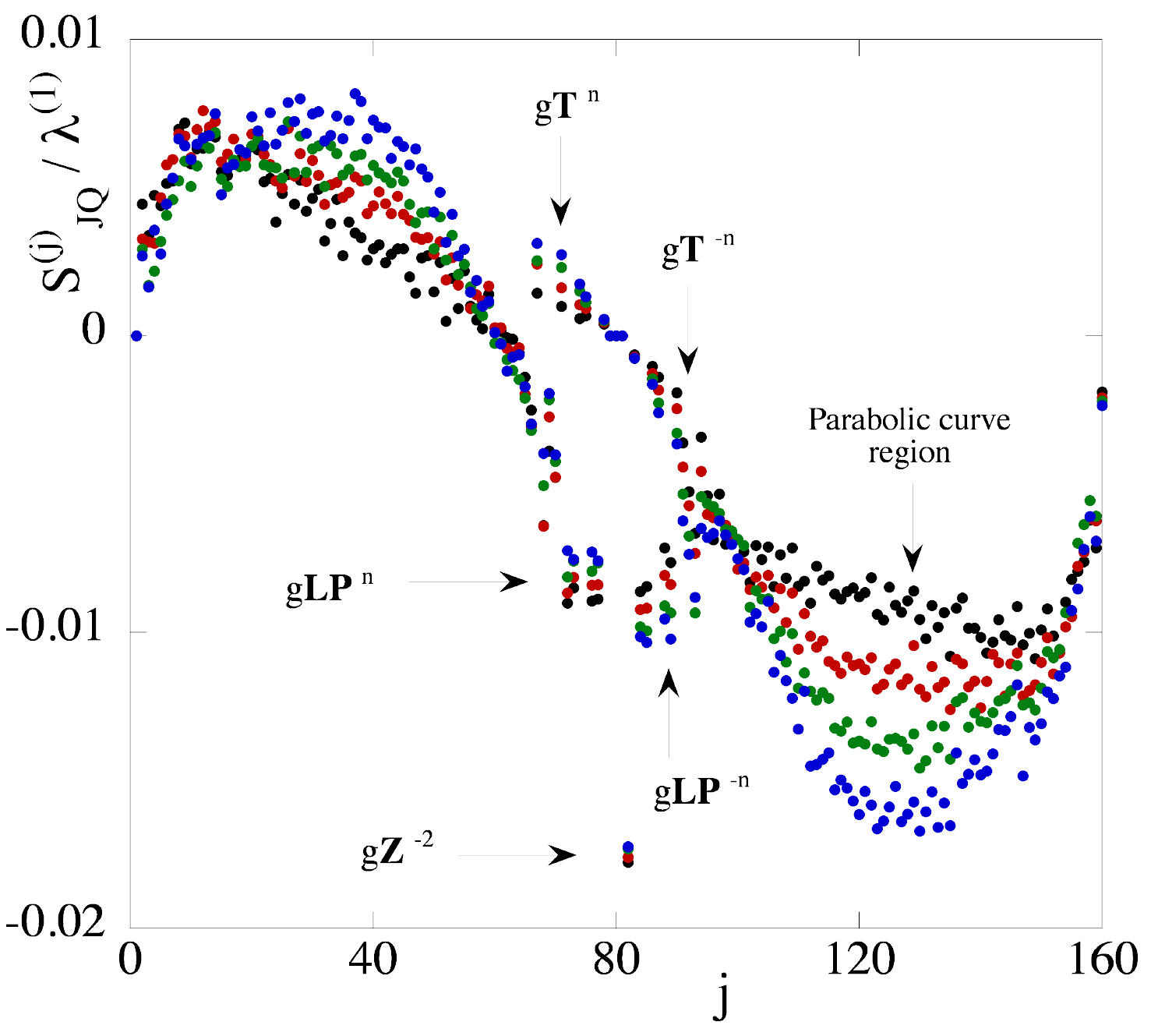}
\caption[Normalised Exponent Spectrum Differences]{Normalised differences between equilibrium and nonequilibrium exponent spectrums $(\lambda^{(j)}_{J_Q}/\lambda^{(1)}_{J_Q}) - (\lambda^{(j)}_{Eq}/\lambda^{(1)}_{Eq}) = S^{(j)}_{J_Q}/\lambda^{(1)}$ for the high density systems, $\rho=0.8$, $N=40$. The differences segment into regions displaying distinct scaling behaviour. Black circles are for $J_Q =0$, red circles for $J_Q =1$, green circles for $J_Q =3$ and blue circles for $J_Q =5$.}
\label{fig:expsC}
\end{figure}

   The normalised high density exponent spectrums of  Fig. \ref{fig:normexpsratio} (b) show small but distinct differences across the spectrum. Fig. \ref{fig:expsC} shows the differences (as opposed to the ratios) between the \emph{normalised} equilibrium and nonequilibrium exponent spectrums $(\lambda^{(j)}_{J_Q}/\lambda^{(1)}_{J_Q}) - (\lambda^{(j)}_{Eq}/\lambda^{(1)}_{Eq}) = S^{(j)}_{J_Q}/\lambda^{(1)}$ (which we have labeled as $S^{(j)}_{J_Q}$ and used the $\lambda^{(1)}$ factor to indicate the normalised values, as $\lambda^{(1)}$ will be different for each system). Fig. \ref{fig:expsC} shows that the exponents within the positive continuous region (the left half of the figure) differ by an approximately consistent positive amount. The exponents in the negative continuous region (the right half of the figure) differ by a negative amount which is not consistent and is \emph{greater} than the positive exponents. This extra negative difference is comparable in magnitude to the \emph{total} positive continuous region exponent difference. Because of this, the negative continuous region exponents in the right half of the difference spectrums appear to form a parabolic curve (indicated by the parabolic label in Fig. \ref{fig:expsC}), which becomes more prominent with increasing heat current.

   Looking within the central mode region of Fig. \ref{fig:expsC} - exponents 60 to 100 - we see two clearly separated branches. The top branch corresponds to the transverse modes (indicated by the $\gT^n$ and $\gT^{-n}$ labels), the bottom branch corresponds to the longitudinal momentum modes (indicated by the $\gLP^n$ and $\gLP^{-n}$ labels). As can be seen, the exponents for the conjugate transverse modes $\gT^n$ and $\gT^{-n}$ differ by an equal and opposite amount, with magnitudes increasing proportional to the increase in heat current.

   Displaying different behaviour from the other vectors, the \emph{positive} longitudinal momentum modes are seen to show a \emph{negative} difference to the normalised equilibrium exponents, opposite to the other positive exponents (and indicated by the $\gLP^n$ label in Fig. \ref{fig:expsC}). This difference is seen to \emph{decrease} in magnitude as the heat current increases. The negative longitudinal momentum exponents also show a comparable negative difference to the positive exponents (indicated by the $\gLP^{-n}$ label), which increases in magnitude with increasing heat current. 
   
   Of the four central zero mode exponents, the three exponents associated with the modes $\gZ^{y}$, $\gZ^{t}$ and $\gZ^{p_y}$ remain zero regardless of the heat current imposed, while the exponent associated with $\gZ^{E}$, $\lambda^{(2N+2)}_{J_Q} = \lambda^{-Z2}_{J_Q}$, shows a large negative difference (indicated by the $\gZ^{-2}$ label in Fig. \ref{fig:expsC}). This is because the mode associated with this exponent, indicating the energy conservation in the system $\gZ^{E}$, is broken for nonequilibrium systems and becomes a nonzero lyapunov mode. The zero mode exponent $\lambda^{(2N+2)}$ shows the largest differences in the spectrum between the normalised equilibrium and nonequilibrium exponents, and again proportional to heat current.

\begin{figure}[ht]
   \includegraphics[width=0.45\textwidth]{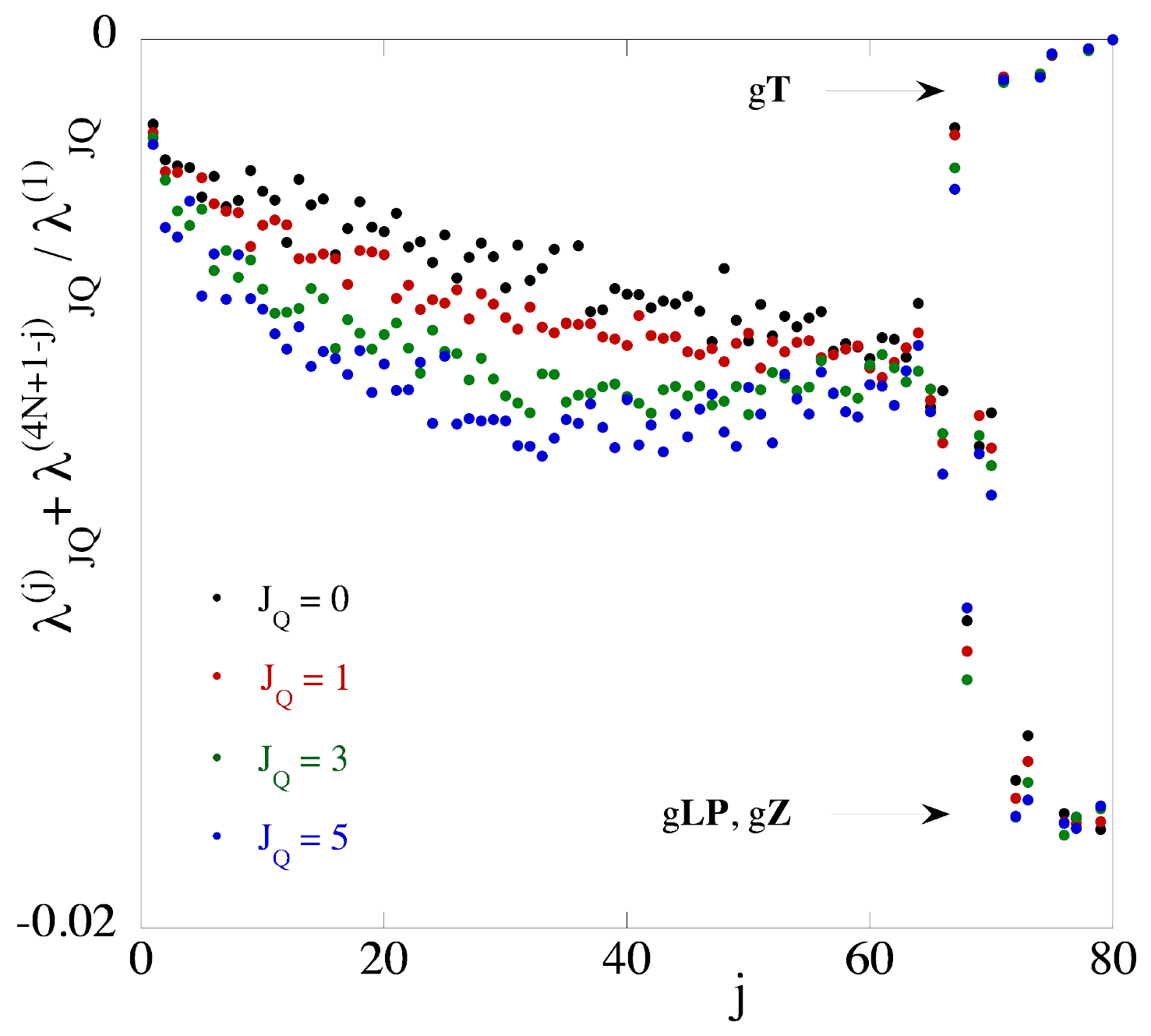}
\caption[Normalised Conjugate Exponent Summation]{Normalised symplectic differences between conjugate exponent pairs for nonequilibrium high density states $(\lambda^{(j)}_{J_Q} + \lambda^{(4N+1-j)}_{J_Q})/\lambda^{(1)}_{J_Q}$ indicating deviations from the conjugate pairing rule for $\rho=0.8$, $N=40$. The symplectic differences between conjugate pairs for the equilibrium system would be exactly zero. Black circles are for $J_Q =0$, red circles for $J_Q =1$, green circles for $J_Q =3$ and blue circles for $J_Q =5$.}
\label{fig:expsD}
\end{figure}

   Fig. \ref{fig:expsD} shows the sum of each individual symplectic pair of normalised conjugate exponent pairs $(\lambda^{(j)}_{J_Q} + \lambda^{(4N+1-j)}_{J_Q})/\lambda^{(1)}_{J_Q}$ for the nonequilibrium high density systems (the symplectic pairing for an equilibrium system is exact and would give a set of $2N$ zeros).  We see that the introduction of a heat current breaks the symplectic pairing condition. The \emph{normalised} symplectic pairing is used to see the properties of the symplectic pairing independent of the uniform magnitude increase due to the heat current (that would be present if $\lambda^{(j)}_{J_Q} + \lambda^{(4N+1-j)}_{J_Q}$ were used directly).

   There are three clear regions to the symplectic pairing of Fig. \ref{fig:expsD}; the main continuous region exponent pairings (for pairs from $j = 1$ to 60), the transverse exponent pairings with zero magnitude (indicated by the $\gT$ label) and the longitudinal momentum mode pairings with approximately twice the magnitude as the continuous region pairings (indicated by the $\gLP$ label). 
   
   All features of the high density spectrums, from Figs. \ref{fig:expratios} (b) to \ref{fig:expsD}, can be explained. The approximately constant negative nonzero summation of the symplectic pairs in the continuous region of the exponent spectrum (from $j = 1$ to 60) in Fig. \ref{fig:expsD} is the additional negative shift of the normalised negative continuous region exponents discussed previously. This additional shift compared to the positive continuous exponents  (of magnitude comparable to the total positive exponent difference) is the cause of the resulting parabolic shape of the normalised negative continuous region differences seen in Fig. \ref{fig:expsC}. 
   As the increase in the conjugate transverse modes' normalised exponents were equal and opposite (seen via the $\gT^n$ and $\gT^{-n}$ labels of Fig. \ref{fig:expsC}), their symplectic differences in Fig. \ref{fig:expsD} gives zero magnitude, regardless of the magnitude of the applied heat current.

\begin{figure*}
                \includegraphics[width=\textwidth]{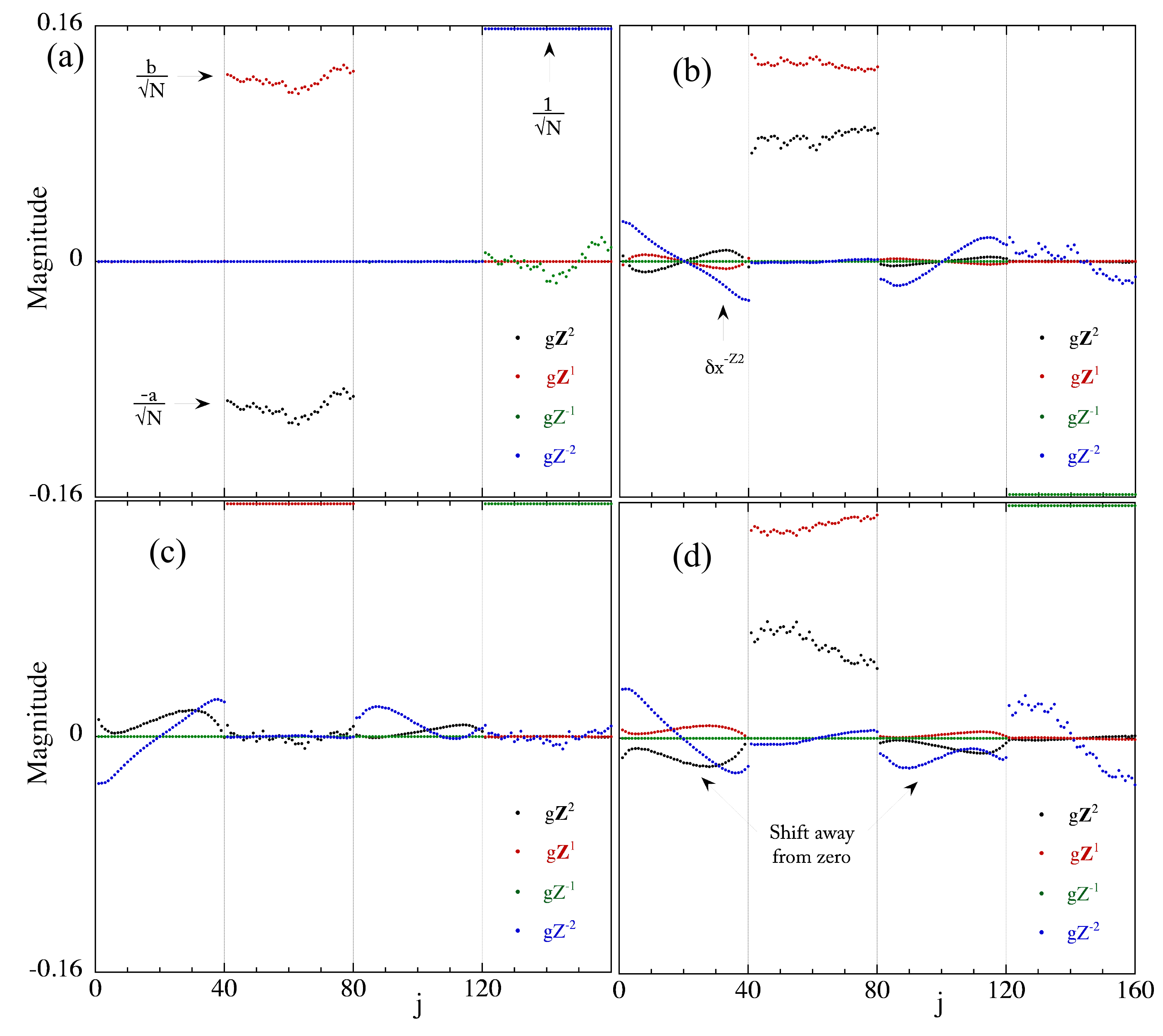}
                \caption{The average zero modes of the high density systems, $\rho=0.8$ and $N=40$ for both equilibrium and three nonequilibrium configurations,  panel (a) for the four \textbf{Equ} zero modes, panel (b) for the four $\mathbf{J_Q=0}$ zero modes,  panel (c) for the four $\mathbf{J_Q=1}$ zero modes and panel (d) for the four $\mathbf{J_Q=3}$ zero modes. As the total momenta is zero (the system does not move), the distribution of the momenta is removed when the modes are averaged. For each nonequilibrium system the exponent of the $\gZ^{-2}$ mode shifts downwards away from zero (shown in Fig. \ref{fig:expsC}). The forms here verifies the nonequilibrium $\delta x$ components witnessed in \cite{TanMor2007a}.}
\label{fig:gsheatzeros}
\end{figure*}

   We saw from the $\gLP^n$ and $\gLP^{-n}$ labels in Fig. \ref{fig:expsC} both conjugate longitudinal momentum mode exponents show a comparable \emph{negative} difference, and Fig. \ref{fig:expsD} shows their relation to the broken zero mode. The symplectic difference for the exponents of the zero mode pair returns a negative shift equal to the broken zero mode exponent by definition ($\lambda^{(2N-1)}_{J_Q}+\lambda^{(2N+2)}_{J_Q} = 0+ \lambda^{(2N+2)}_{J_Q}$), which we can label as $S^{Z}_{J_Q}$ (as these are normalised this should technically be $S^{Z}_{J_Q}/\lambda^{(1)}_{J_Q}$). We see from Fig. \ref{fig:expsD} the symplectic difference for the longitudinal momentum modes is seen to be equal to the zero mode shift amount  $S^{Z}_{J_Q}$ (indicted in the figure by the $\gLP, \gZ$ label). This means as well as the increase from the heat current $R_{J_Q}$, each $\gLP$ mode carries \emph{half} the shift of the broken zero mode shift $ S^{LP}_{J_Q} = 1/2 \; S^{Z}_{J_Q}$ for all nonequilibrium systems. \emph{Both} the positive and negative longitudinal momentum mode shifts downwards by an amount approximately equal to the negative continuous region difference (the $\gZ^{-2}$ mode twice as much) resulting in the zero and longitudinal momentum symplectic pairs having equal, and approximately twice, the symplectic difference as the continuous region exponents.

   The shift downwards of each $\gLP$ mode, equal to half the $\gZ^{-2}$ shift, is a constant amount independent of the mode number. By fitting the broken zero mode shift (equal to the exponent $\lambda^{-Z2}_{J_Q}$), a power law dependency of the shift as a function of the heat current can be found for the $\gLP$ modes
\begin{equation}\label{equ:slpshift}
   S^{LP}_{J_Q} = \lambda^{-Z2}_{J_Q}/2 \cong -0.038 -0.018 (J_Q)^{1/2}
\end{equation}
   for the true exponent values (not the normalised exponents), independent of the mode number. 
   Using this with our previous expression, the Lyapunov exponent spectrum of a nonequilibrium system with any heat current is given by
\begin{eqnarray}
\label{equ:heatexpsratio}   \lambda^{(j)}_{J_Q} &=& R_{J_Q}  \lambda^{(j)}_{Eq} \\
\label{equ:heatexpsratio2}   \lambda^{LP}_{J_Q} &=& R_{J_Q}  \lambda^{LP}_{Eq} + S^{LP}_{J_Q}. \\
\label{equ:heatexpsratio3}   \lambda^{-Z2}_{J_Q} &=& 2 \; S^{LP}_{J_Q} 
\end{eqnarray}
%

\begin{figure*}
                \includegraphics[width=\textwidth]{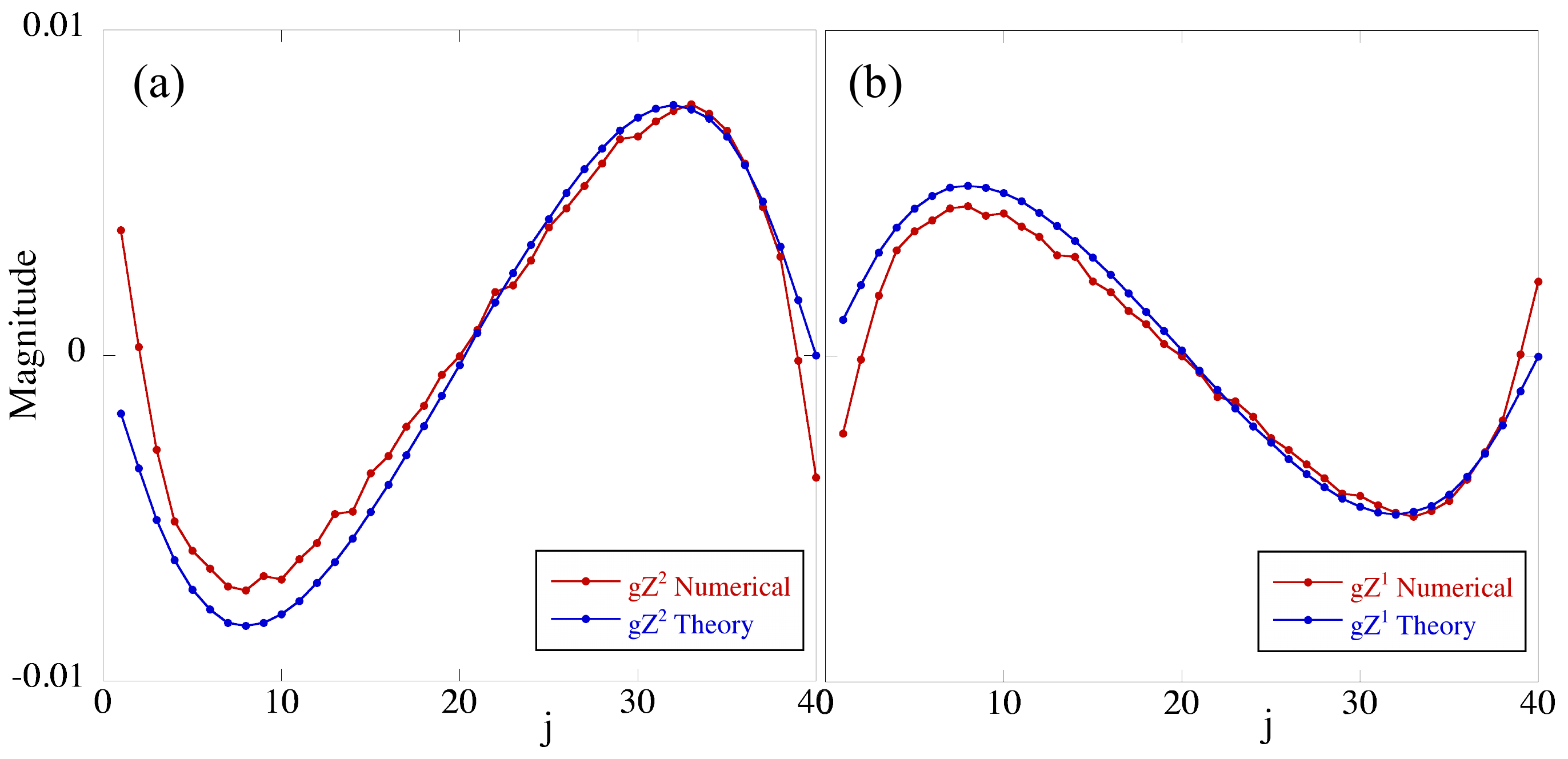}
                \caption{A comparison between the numerical $\delta x$ components and the predicted $\delta x$ components of the average $\gZ^{2}$ and $\gZ^{1}$ zero modes as given by Eqs. (\ref{equ:heatz2delx}) and (\ref{equ:heatz1delx}) for the $J_Q=0$, $\rho=0.8$, $N=40$ system of Fig. \ref{fig:gsheatzeros} (b). panel (a) for $\delta x$ for the $\gZ^{2}$ mode and panel (b) for $\delta x$ for the $\gZ^{1}$ mode.}
\label{fig:deltaxwalls}
\end{figure*}

   Eq. (\ref{equ:heatexpsratio}) shows the majority of the exponents are given by a direct ratio $R_{J_Q}$ ($R_{J_Q}$ given by Eq. (\ref{equ:rjratio})), while Eq. (\ref{equ:heatexpsratio2}) shows the positive and negative $\gLP$ modes are given by the ratio and an addition constant negative shift (the constant shift given by Eq. (\ref{equ:slpshift})). This also explains the $\gLP$ ratio effect in Fig. \ref{fig:expratios} (b). The shift $S^{LP}_{J_Q}$ is constant, therefore the first LP mode - having the smallest magnitude - is affected the most and shows the largest shift away from the expected $R_{J_Q}  \lambda^{LP}_{Eq}$ value, and explains the points indicated by the $\gLP^n$ labels in Fig. \ref{fig:normexpsratio} (b). As the magnitude of the $\gLP$ exponent increases with increasing mode number the effect the constant shift has in comparison to the direct scaling decreases. 
   
   Fig. \ref{fig:expsD} also shows the effect mode mixing has on the behaviour of the modes, as vectors $g^{(68)}$, $g^{(69)}$ and $g^{(70)}$ are $\gT$-$\gLP$ mixed mode vectors. The $\gLP$ downwards shift $ S^{LP}_{J_Q}$ is shared amongst these modes, resulting in the positive transverse mode shifting downwards as opposed to upwards, while the longitudinal momentum modes are shifted up slightly (seen above the $\gLP^n$ modes label in Fig. \ref{fig:expsC}). When the conjugate pairs are added together this results in the longitudinal momentum pair only giving half the normal shift, while the transverse mode negative shift becomes much larger (its conjugate also shifts downwards). This gives the single point seen between the $\gLP$ conjugate pairings and the continuous region pairings, as well as the single $\gLP$ pair with pairing magnitude close to the continuous region pairing magnitude.

\section{Non-Equilibrium Backwards Lyapunov Modes}\label{subsec:NEBLV}

\subsection{Zero modes}\label{subsec:heatzeromodes}

   As the BLV zero modes form a segregated subspace of the tangent space (at equilibrium) we can analyse their change independent of the other vectors. The total momenta of the system is always zero $\left< \mathbf{p}_i \right> = 0$ (for both equilibrium and nonequilibrium systems) therefore the components of the zero modes proportional to the instantaneous momenta, when averaged, will sum to zero. By averaging the modes only the components not related to the instantaneous momenta will remain and allows the structure of the modes to be seen. Fig. \ref{fig:gsheatzeros} shows all four averaged zero modes for high density nonequilibrium systems with heat currents $J_Q=0$, 1 and 3 (Figs. \ref{fig:gsheatzeros} (b), \ref{fig:gsheatzeros} (c) and \ref{fig:gsheatzeros} (d) respectively) compared to the equilibrium system (Fig. \ref{fig:gsheatzeros} (a)). 
      
   Consistent with Section \ref{equilanal}, each element of the averaged equilibrium $\gZ^{2}$ mode of Fig. \ref{fig:gsheatzeros} (a) contains only the $\gZ^{y}$ shift factored by an amount $a < 1$, while the $\gZ^{1}$ mode contains the same shift from $\gZ^{y}$ factored by $b$. This gives
\begin{eqnarray*}
   \left< \gZ^{2}_i \right> &=& -a \left< \gZ^{y}_i \right> + b \left< \gZ^{t}_i \right> = -\frac{a}{\sqrt{N}} \\
   \left< \gZ^{1}_i \right> &=& b \left< \gZ^{y}_i \right> + a \left< \gZ^{t}_i \right> = \frac{b}{\sqrt{N}}
\end{eqnarray*}
    in the $\delta y$ components of the zero modes as the $\gZ^{n}_i \propto \mathbf{p}_{i}$ dependence is removed on average and is indicated by the $-a/\sqrt{N}$ and $b/\sqrt{N}$ labels in Fig. \ref{fig:gsheatzeros} (a).
   The equilibrium $\gZ^{-1}$ mode, comprised solely of the $\gZ^{E}$ basis vector, averages to zero with small fluctuations. The equilibrium $\gZ^{-2}$ mode, comprised solely of the $\gZ^{p_y}$ shift equal to $1/\sqrt{40} = 0.158$, equals the largest \emph{average} value obtainable for each of the zero modes and is indicated by the $1/\sqrt{N}$ label in Fig. \ref{fig:gsheatzeros} (a).

\begin{figure*}
                \includegraphics[width=\textwidth]{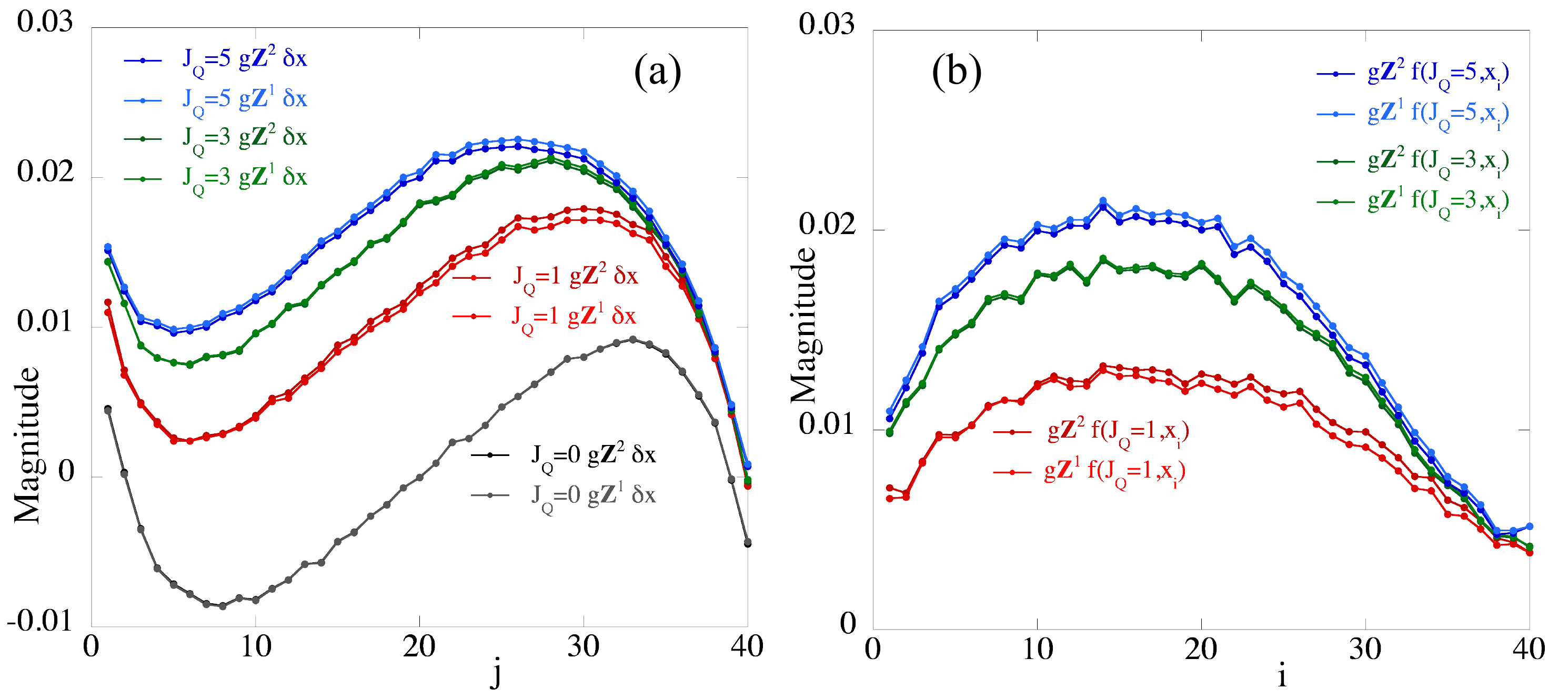}
                \caption{Panel (a) shows the $\delta x$ components for all nonequilibrium $\gZ^{2}$ and $\gZ^{1}$ modes, showing the exact average functional form as given by Eqs. (\ref{equ:heatz2delx}) and (\ref{equ:heatz1delx}) when the $a$ and $b$ factors are removed. Panel (b) shows the differences between the $J_Q\neq0$ and $J_Q=0$ $\delta x$ components, giving the $f(J_Q,x_i)$ function.}
\label{fig:deltaxs}
\end{figure*}

   For the nonequilibrium systems the $\gZ^{-1}$ mode, which previously corresponded to the (now broken) energy conservation mode, swaps position with the $\gZ^{-2}$ mode (as the modes are organised in the order of their exponents magnitude and $ \lambda^{-Z2}_{J_Q} = 2 \; S^{LP}_{J_Q} < 0$). Therefore the nonequilibrium $\gZ^{-1}$ mode now corresponds to the straight shift $\gZ^{p_y}$ basis vector, while the nonequilibrium $\gZ^{-2}$ mode is now associated with the $\gZ^{E}$ basis vector. 
   
   For the $J_Q=0$ system (Fig. \ref{fig:gsheatzeros} (b)) the $\gZ^{-2}$ mode (in blue) contains a clear linear function across the system in the $\delta x$ component and an almost full sine function in the $\delta p_x$ component (the $\delta x$ component is indicated by the $\delta x^{-Z2}$ label in Fig. \ref{fig:gsheatzeros} (b)); a verification of the $\delta x$ components witnessed in \cite{TanMor2007a}. The $\gZ^{2}$ mode (in black) and the $\gZ^{1}$ mode (in red) are seen to have similar sine functions in their $\delta x$ components as the $\gZ^{-2}$ mode's $\delta p_x$ component.
   The $\delta x$ components of the $\gZ^{2}$ and $\gZ^{1}$ modes can be described very accurately by altering their functional forms to
\begin{eqnarray}
\label{equ:heatz2delx}    \delta x^{Z2}_i &=&  - b p_{xi} \rightarrow -b \left(p_{xi} + \delta x^{-Z2}_i \sin{\frac{\pi x_i}{L_x}} \right) \\
\label{equ:heatz1delx}    \delta x^{Z1}_i &=&  a p_{xi} \rightarrow a \left(p_{xi} + \delta x^{-Z2}_i \sin{\frac{\pi x_i}{L_x}} \right)
\end{eqnarray}
   where the $a=0.532$ and $b=0.846$ values can be found from the $\delta y$ components of the modes. This implies the $\delta x$ components of the $\gZ^{2}$ and $\gZ^{1}$ modes in the $J_Q=0$ system contains the product of the linear form of the $\gZ^{-2}$ $\delta x$ component and the first equilibrium $\gL$ mode functional form. Fig. \ref{fig:deltaxwalls} shows a comparison between this predicted functional form and the numerically calculated average $\delta x$ for both positive zero modes; Fig. \ref{fig:deltaxwalls} (a) for the $\gZ^{2}$ mode and Fig. \ref{fig:deltaxwalls} (b) for the $\gZ^{1}$ mode.

   From such a relatively simple alteration to the functional form very close agreement can be seen. The functional form $\delta x^{Z2}_i \sim \delta x^{-Z2}_i \sin{\pi x_i/L_x} $ has to be used instead of $\delta x^{Z2}_i \sim \sin{2\pi x_i/L_x} $ as the maximal values would no longer be coincident, and would also imply the zero mode adopts the $\gL^2$ functional form over $\gL^1$, which appears unlikely. The most apparent explanation for this is to ensure the zero modes remain orthogonal to each other.

   The zero modes for the $J_Q\neq0$ systems in Figs. \ref{fig:gsheatzeros} (c) and \ref{fig:gsheatzeros} (d) have similar functional forms to the $J_Q=0$ system. While the same basic features in the average components can be seen for all the nonequilibrium systems, the $a$ and $b$ values change randomly. A constant shift is also seen in the $\delta x$ and $\delta p_x$ components of the $\gZ^{2}$ and $\gZ^{1}$ modes and the $\delta p_x$ components of the $\gZ^{-2}$ mode when $J_Q\neq0$ (indicated by the label in Fig. \ref{fig:gsheatzeros} (d)).
   This shift is seen more clearly in Fig. \ref{fig:deltaxs} (a), which shows the $\delta x$ components of the $\gZ^{2}$ and $\gZ^{1}$ modes for all nonequilibrium systems with their respective $a$ and $b$ leading factors removed. The $\delta x$ components for the $J_Q=0$ system (as given in Fig.s \ref{fig:deltaxwalls} (a) and \ref{fig:deltaxwalls} (b) with their respective $a$ and $b$ factors) are the lowest two functions of Fig. \ref{fig:deltaxs} (a). With the removal of the $a$ and $b$ factors both $\delta x^{Z1}_i$ and $\delta x^{Z2}_i$ are seen to have the exact same functional form, which is to be expected.
   
   The most intriguing aspect of Fig. \ref{fig:deltaxs} (a) is the clear dependence on the magnitude of the heat current $J_Q$ in the movement of the $\delta x$ functional form away from the symmetric $J_Q=0$ case. Proportional to the heat current, the $\delta x$ functional form shifts asymmetrically upwards. This implies that the $\delta x$ components for the systems with a nonzero heat current is given by, without the $a$ or $b$ factor
\begin{eqnarray*}
    \delta x^{Zn}_i &=& \left(p_{xi} + \delta x^{-Z2}_i \sin{\frac{\pi x_i}{L_x}} \right) \\
    &\rightarrow& \left(p_{xi} + \delta x^{-Z2}_i \sin{\frac{\pi x_i}{L_x}} + f(J_Q,x_i) \right).
\end{eqnarray*}
%

\begin{figure*}
                \includegraphics[width=\textwidth]{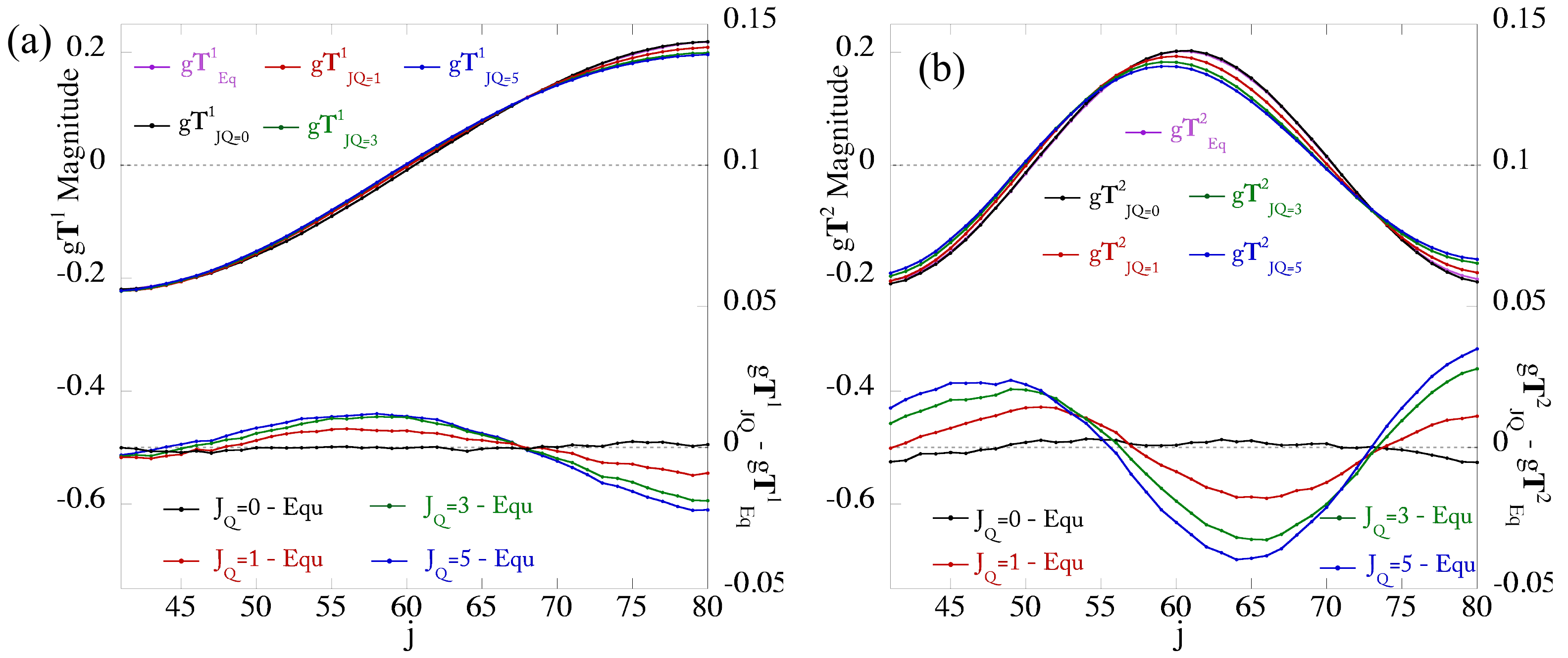}
                \caption{The $\delta y$ components of the nonequilibrium backward transverse modes in comparison to the equilibrium transverse mode, for $\rho=0.8$ and $N=40$, panel (a) for the  $\gT^1$ modes, panel (b) for the  $\gT^2$ modes. The top of each figure shows the transverse modes themselves (magnitude given by the left hand axis), the bottom gives the (enlarged) differences between the nonequilibrium and equilibrium transverse modes (magnitude given by the right hand axis).}
\label{fig:heattmodeschage}
\end{figure*}

   What the function $f(J_Q,x_i)$ is can be seen in Fig. \ref{fig:deltaxs} (b). Taking the difference between the $J_Q\neq0$ and $J_Q=0$ $\delta x$ components shows that the $f(J_Q,x_i)$ function takes the form of a skewed sine function $f(J_Q,x_i) = c_{J_Q} \sin{\pi x_i/L_x} + d_{J_Q}$ (similar to the $J_Q=0$ augmentation), where $c_{J_Q}$ and $d_{J_Q}$ are dependent on the magnitude of the heat current.
   
   Independent of the magnitude of the heat current, there is a clear relationship between the numerical $\gZ^{2}$ and $\gZ^1$ modes. The ratio of the modes $\delta x$ and $ \delta y$ components is numerically seen to be
\begin{eqnarray*}
   \frac{\delta x^{Z2}}{\delta x^{Z1}} = \frac{\delta p_x^{Z2}}{\delta p_x^{Z1}} = \frac{a}{b} \; \; \; \; \; \; \; \; \mathrm{and} \; \; \; \;  \; \; \; \; \frac{\delta y^{Z2}}{\delta y^{Z1}} = \frac{b}{a}
\end{eqnarray*}
   with exact agreement pointwise across all elements. As the sign of $a$ and $b$ can change arbitrarily for each system, one of these ratios will be negative. This is encouraging and implies the functional forms for both the $\gZ^2$ and $\gZ^1$ modes are described by the same functional form, simply with different leading constant factors. The $\delta p_y$ components of the modes do not seem to mirror the $\delta y$ relationship, but instead follow the $\delta x$ and $\delta p_x$ relation. This is thought to be due to the fact the perturbation away from equilibrium is along the $x$ direction, therefore the components changed (whether they be in $x$ or $y$) will be proportional to the $x$ components. 

   There is a known relationship between the coordinate and momenta components of each Lyapunov vector; $\delta p^{(j)} = \lambda^{(j)}\delta q^{(j)} $ \cite{ChuTru2010}. At equilibrium this relationship technically still held for the positive zero modes, as $\delta p = 0$ components were zero everywhere (and vice versa for the negative zero modes). 
   For nonequilibrium systems the $\delta p_x$ and $\delta x$ components of the $\gZ^2$ and $\gZ^1$ modes show an intriguing relationship. The relationship between the components seem to follow the reverse of the temperature profile of the system
\begin{eqnarray*}
   \delta p_x^{Zn} = c \: T_{N-x} \: \delta x^{Zn}.
\end{eqnarray*}
   The $c$ constant is required as the ratio follows the shape, but not the exact values, of the temperature profile (as given in Fig. \ref{fig:tempprof} (b)). This ratio is seen to be the same pointwise for both the $\gZ^2$ and $\gZ^1$ modes leading to the conclusion that it is not coincidental, although the explanation for it remains to be found.

\subsection{Transverse and Longitudinal Momentum Modes}\label{subsec:heattrans}

   We saw from Fig. \ref{fig:expsC} that with the introduction of a heat current the exponents related to the stationary transverse modes increase in magnitude substantially (given via Eq. (\ref{equ:heatexpsratio})). As the heat current substantially changed the form of the backward zero modes, we can also see what effect the heat current has on the functional form of the transverse modes.

   Fig. \ref{fig:heattmodeschage} shows the $\delta y$ components of the (time-averaged) nonequilibrium transverse modes compared to the equilibrium transverse mode, the top of Fig. \ref{fig:heattmodeschage} (a) for the $\gT^1$ modes and the top of Fig. \ref{fig:heattmodeschage} (b) for the $\gT^2$ modes. The bottom of Figs. \ref{fig:heattmodeschage} (a) and \ref{fig:heattmodeschage} (b) shows the differences between the nonequilibrium and equilibrium transverse modes. It is important to note the scale for the differences between the nonequilibrium and equilibrium modes (magnitude given by the right hand axis) has been enlarged considerably in order to see the small differences .
   
   From Fig. \ref{fig:heattmodeschage} we see that the nonequilibrium transverse modes show a small change in the functional form due to the heat current, given approximately as
\begin{eqnarray*}
   \delta y^{Tn}_{J_Q} - \delta y^{Tn}_{Eq} \sim \cos{\left[(n+\frac{1}{2})\frac{\pi x_i}{L_x}\right]}
\end{eqnarray*}
   which increases in magnitude with increasing heat current. This shows that while the magnitude of the nonequilibrium transverse mode exponent is increased significantly, the functional form of the transverse modes are altered only very slightly by the application of a heat current onto the system. This increase in the magnitude of the exponent also increases the relative size of the $\delta p_y$ components of the nonequilibrium transverse modes. This is due to the relationship between the coordinate and momentum components of the vector $\delta p^{(j)} = \lambda^{(j)}\delta q^{(j)} $, as the magnitude of $\lambda^T_n$ increases so does the relative size of $\delta p^T_y$. 
   
   The effect the heat current has on the longitudinal momentum modes will be discussed in Section \ref{subsec:NEAS}.
   
\section{Non-Equilibrium Covariant Lyapunov Modes}\label{subsec:NECLV}

   Like for the equilibrium systems, the exponent of the $j$th nonequilibrium covariant vector is identical to the exponent of the $j$th nonequilibrium backward vector. As the CLVs are not constrained by orthogonality, this significantly alters their nonequilibrium forms compared to the BLVs.

\begin{figure}[ht]
   \includegraphics[width=0.45\textwidth]{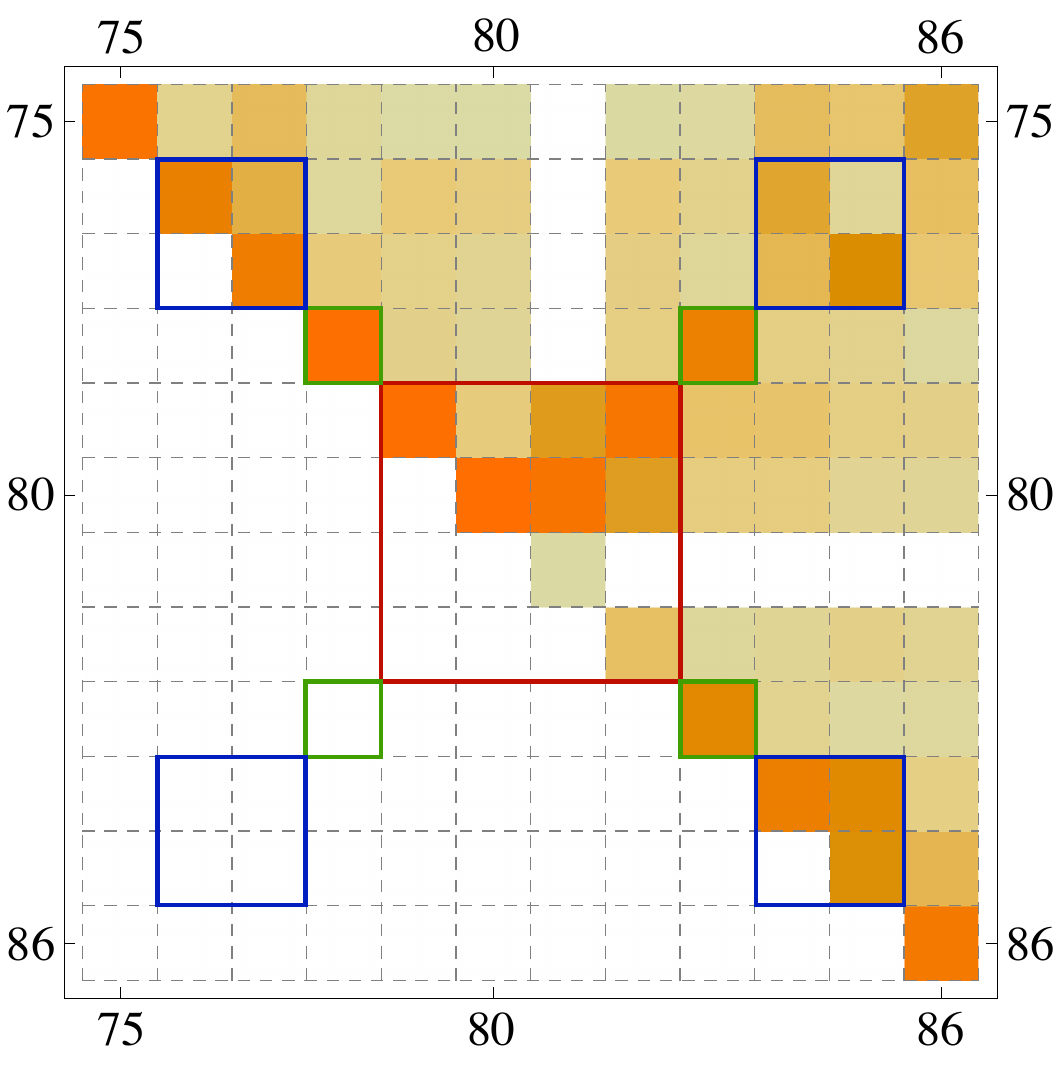}
\caption[Nonequilibrium $C$ Matrix Centre]{The centre of an evolved nonequilibrium $C$ matrix corresponding to the $N=40$, $\rho=0.8$, $J_Q=3$ system. Shown over the same region as the equilibrium $C$ matrix of Fig. \ref{fig:Cidentity} and utilising the same mode identification colour boxes, there are clear differences between the equilibrium and nonequilibrium matrices.}
\label{fig:heatcmatrix}
\end{figure}

   Fig. \ref{fig:heatcmatrix} shows the central mode region of an indicative evolved $C$ matrix for the nonequilibrium $N=40$, $\rho=0.8$, $J_Q=3$ system. Shown over the same central region as the equilibrium $C$ matrix of Fig. \ref{fig:Cidentity}, there are clear differences between the equilibrium and nonequilibrium matrices. 

   In contrast to the equilibrium covariant zero modes, the nonequilibrium covariant zero modes show significant interaction with the covariant hydrodynamic Lyapunov modes.
  Compared to the clear zero bands in the equilibrium $C$ matrix of Fig. \ref{fig:Cidentity} - which indicates the segregation of the covariant zero modes - the $C$ matrix entries corresponding to the $\vZ^1$, $\vZ^2$ and $\vZ^{-2}$ modes all contain nonzero values above and to the right of the central $4 \times 4$ subspace (outlined within the same red square as for the equilibrium zero modes). The only zero mode which remains invariant and segregated from all other non zero modes (except for the shift in position along the spectrum) is the $\vZ^{-1}$ mode.
  The transverse and longitudinal momentum modes remain approximately segregated but show increased interaction, discussed below.
     
\subsection{Zero Modes}\label{subsec:cvzeroheats}

\begin{figure*}
                \includegraphics[width=0.9\textwidth]{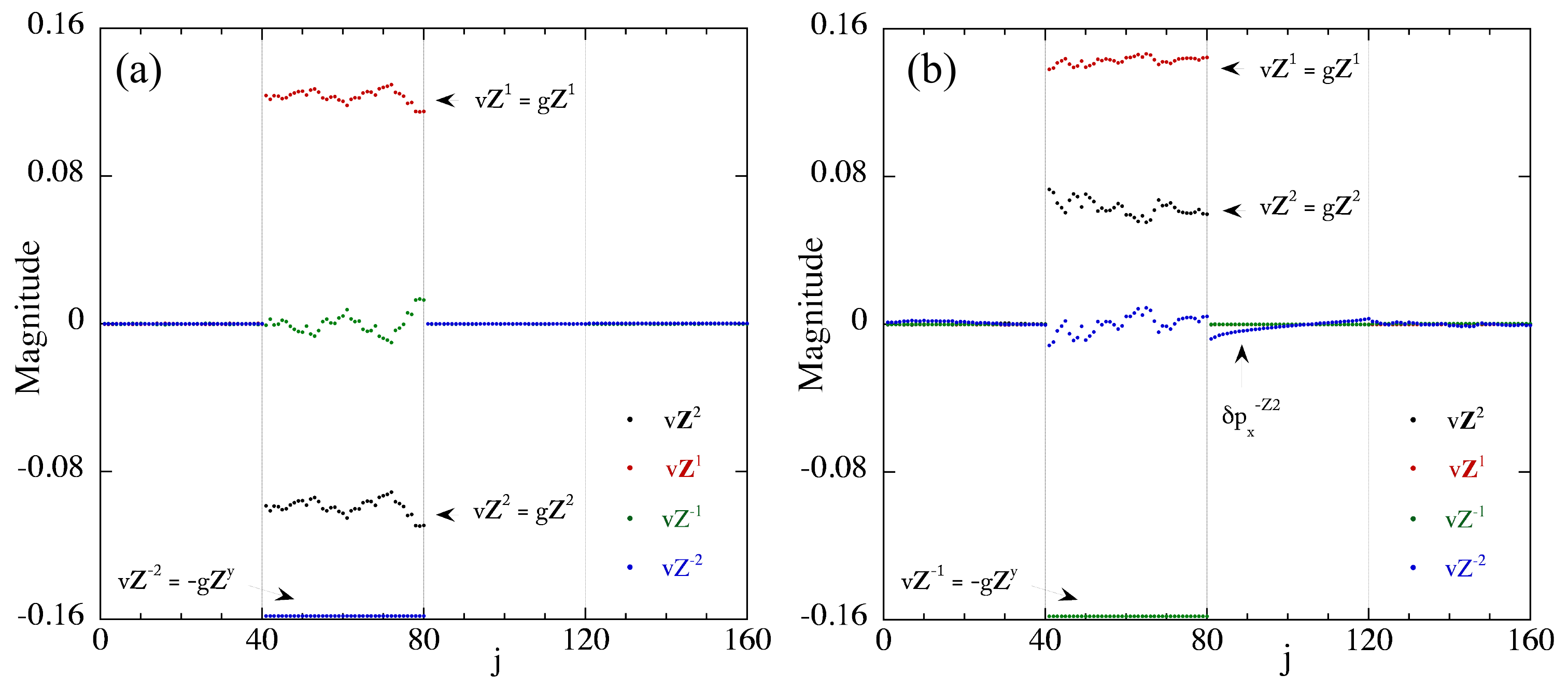}
                \caption{The average four covariant zero modes of equilibrium and nonequilibrium $\rho=0.8$, $N=40$ systems,  panel (a) for the four \textbf{Equ} zero modes, panel (b) for the four $\mathbf{J_Q=3}$ zero modes. Compare directly with the BLVs of Figs. \ref{fig:gsheatzeros} (a) and \ref{fig:gsheatzeros} (d). The centre $C$ matrix of Fig. \ref{fig:heatcmatrix} gives the covariant zero modes of panel (b).}
\label{fig:cvheatzeros}
\end{figure*}

   Section \ref{equilanal} showed the equilibrium positive covariant zero modes are equal to the positive backward zero modes and the negative covariant zero modes are the negative of the $\delta q$ zero mode basis vectors (with small contributions in the $\delta p$ zero mode basis vectors). This can be seen in the labels of Fig. \ref{fig:cvheatzeros} (a), which gives the average covariant zero modes for the equilibrium system. Comparing Fig. \ref{fig:cvheatzeros} (a) to Fig. \ref{fig:gsheatzeros} (a) (the average backwards zero modes for the same equilibrium system) shows the relation between BLVs and CLVs.
   
   Fig. \ref{fig:cvheatzeros} (b) shows the average covariant zero modes for the nonequilibrium system with a heat current of $J_Q = 3$, in comparison to the backwards zero modes as given in Fig. \ref{fig:gsheatzeros} (d) for the same system. The functional forms of the covariant zero modes are radically different; the large contributions in the $\delta x$, $\delta p_x$ and $\delta p_y$ components are essentially removed for the nonequilibrium covariant modes. Analogous to the equilibrium system, the covariant $\vZ^2$ and $\vZ^1$ modes are the same as the backwards zero modes and the $\vZ^{-1}$ mode again is given by $-\gZ^{y}$, equivalent to the equilibrium system (indicated by the labels in Fig. \ref{fig:cvheatzeros} (b)). 
   
   The $\vZ^{-2}$ mode, the source of the large changes in the backward zero modes, has a much simpler covariant structure in the absence of the orthogonality requirement. There is a small average linear function in $\delta p_x$ which can be seen from the $\delta p_x^{-Z2}$ label in Fig. \ref{fig:cvheatzeros} (b). Once this is accounted for, the $\vZ^{-2}$ mode is given exactly as 
\begin{eqnarray*}
\vZ^{-2} = -c_1 \gZ^{t} + c_2 \gZ^{E},
\end{eqnarray*}
   where $c_1=0.9908$ and $c_2=0.1331$ for the $J_Q=3$ system. The $c_1$ and $c_2$ factors do not appear to connect to the positive zero mode $a$ and $b$ values, and $c_2$ is much larger than the $\epsilon$ factor from of the long time convergence of the mode (from the equilibrium covariant evolution of Section \ref{equilanal}). The average $\delta p_x$ linear component of $\vZ^{-2}$ is similar in form to the average $\delta x$ component of $\gZ^{-2}$. This structure is seen for the covariant zero modes of all nonequilibrium systems.

   As the nonequilibrium covariant zero modes $\vZ^{2}$, $\vZ^{1}$ and $\vZ^{-1}$ remain invariant with the application of a heat current, and the broken energy conservation zero mode $\vZ^{-2}$ takes a considerably simple form, this leads to the conclusion that the complex forms seen for the nonequilibrium backward zero modes are mainly formed as a requirement that they remain orthogonal. From Fig. \ref{fig:heatcmatrix} we see the nonequilibrium backwards zero modes interact with the hydrodynamic Lyapunov modes in order to remove the functional forms imposed on them by the Benettin scheme and return the nonequilibrium covariant zero modes to essentially their equilibrium forms.

\subsection{Transverse and Longitudinal Momentum Modes}\label{subsec:heattmodescv}

\begin{figure*}
                \includegraphics[width=\textwidth]{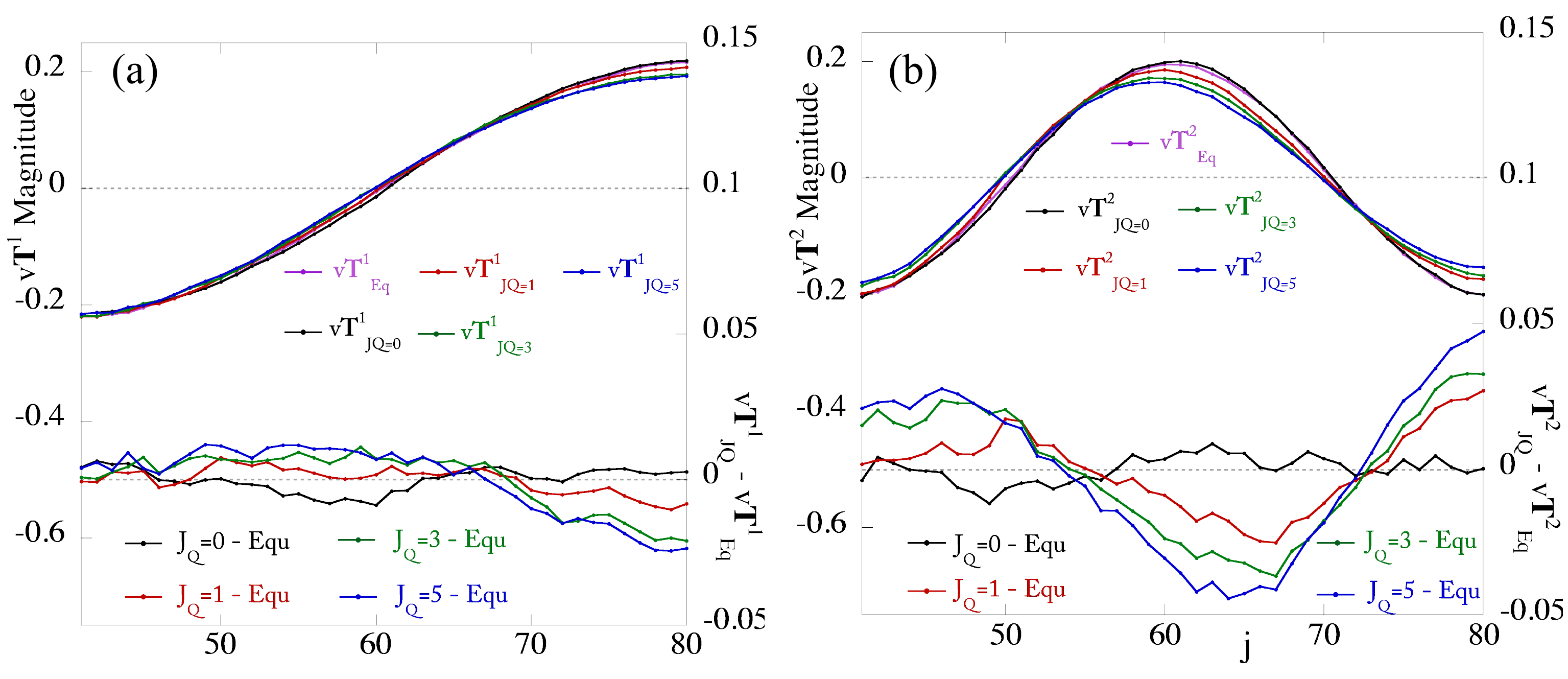}
                \caption{The $\delta y$ components of the nonequilibrium covariant transverse modes in comparison to the equilibrium covariant transverse mode, for $\rho=0.8$ and $N=40$, panel (a) for the  $\vT^1$ modes, panel (b) for the  $\vT^2$ modes. The top of each panel shows the transverse modes themselves (magnitude given by the left axis), the bottom gives the (enlarged) differences between the nonequilibrium and equilibrium transverse modes (magnitude given by the right axis).}
\label{fig:heattmodesCV}
\end{figure*}

   Equal to the backward transverse modes, the exponents of the stationary covariant transverse modes are increased uniformly by the introduction of a heat current. Similar to Fig. \ref{fig:heattmodeschage}, Fig. \ref{fig:heattmodesCV} shows the $\delta y$ components of the nonequilibrium covariant transverse modes compared to the equilibrium transverse modes, the top of Fig. \ref{fig:heattmodesCV} (a) for the $\vT^1$ modes and the top of Fig. \ref{fig:heattmodesCV} (b) for the $\vT^2$ modes. The bottom of Figs. \ref{fig:heattmodesCV} (a) and \ref{fig:heattmodesCV} (b) shows the differences between the nonequilibrium and equilibrium covariant transverse modes. Again it is important to note the difference scale has been enlarged considerably (magnitude given by the right hand axis) in order to see the small differences between the nonequilibrium and equilibrium modes.

   We see from Fig. \ref{fig:heattmodesCV} that the covariant transverse modes change in a similar way to the backward transverse modes. While the same essential differences are seen between the covariant nonequilibrium and equilibrium transverse modes as for the backward transverse modes - and the covariant modes themselves maintain smooth functional forms - there is substantial fluctuation in the difference between the modes. These fluctuations compared to the backward mode differences (even though both the backwards and covariant modes were plotted at the same point in the dynamics) indicates that the removal of the orthogonality condition must allow some of the covariant functional form to spread into other vectors. The direct influence from the heat current seen between the orthogonal backward transverse mode functional forms may not be as strong for the covariant modes.

   The effect the heat current has on the covariant longitudinal momentum modes will be discussed in Section \ref{subsec:NEAS}.
   
\section{Properties of the Lyapunov Modes out of Equilibrium}\label{sec:PotLMooE}

\subsection{Localisations}\label{subsec:NEL}

   A contribution parameter from particle $i$ to each of the Lyapunov vectors $g^{(j)}$ or $v^{(j)}$, $\chi^{(j)}_i = (\delta x_i)^2 + (\delta y_i)^2 + (\delta p_{xi})^2 + (\delta p_{yi})^2$, can be summed over all particles via an entropy function $\sum^{N}_{i=1} \chi^{(j)}_i \ln{\chi^{(j)}_i}$ to form an instantaneous localisation measure
\begin{equation*}
L^{(j)}(t) = \frac{1}{N} \mathrm{exp} \left(-\sum^{N}_{i=1} \chi^{(j)}_i \ln{\chi^{(j)}_i} \right)
\end{equation*}
   This measure, defined between $L^{(j)}(t) = [0,1]$ is close to 0 for a highly localised vector and close to 1 for a highly delocalised vector. A distribution of the instantaneous values is collated to form the average localisation for each vector $\left< L^{(j)} \right>$. Due to the conjugate pairing rule, the symplectic vectors must have the same localisation values.
   
    The thermodynamic limit of the average localisation measures is essentially met for $N=40$, similar to the Lyapunov spectrum. Therefore the average localisation for the equilibrium system has been analysed in detail previously \cite{Morr2012}. 
   
\begin{figure*}
                \includegraphics[width=\textwidth]{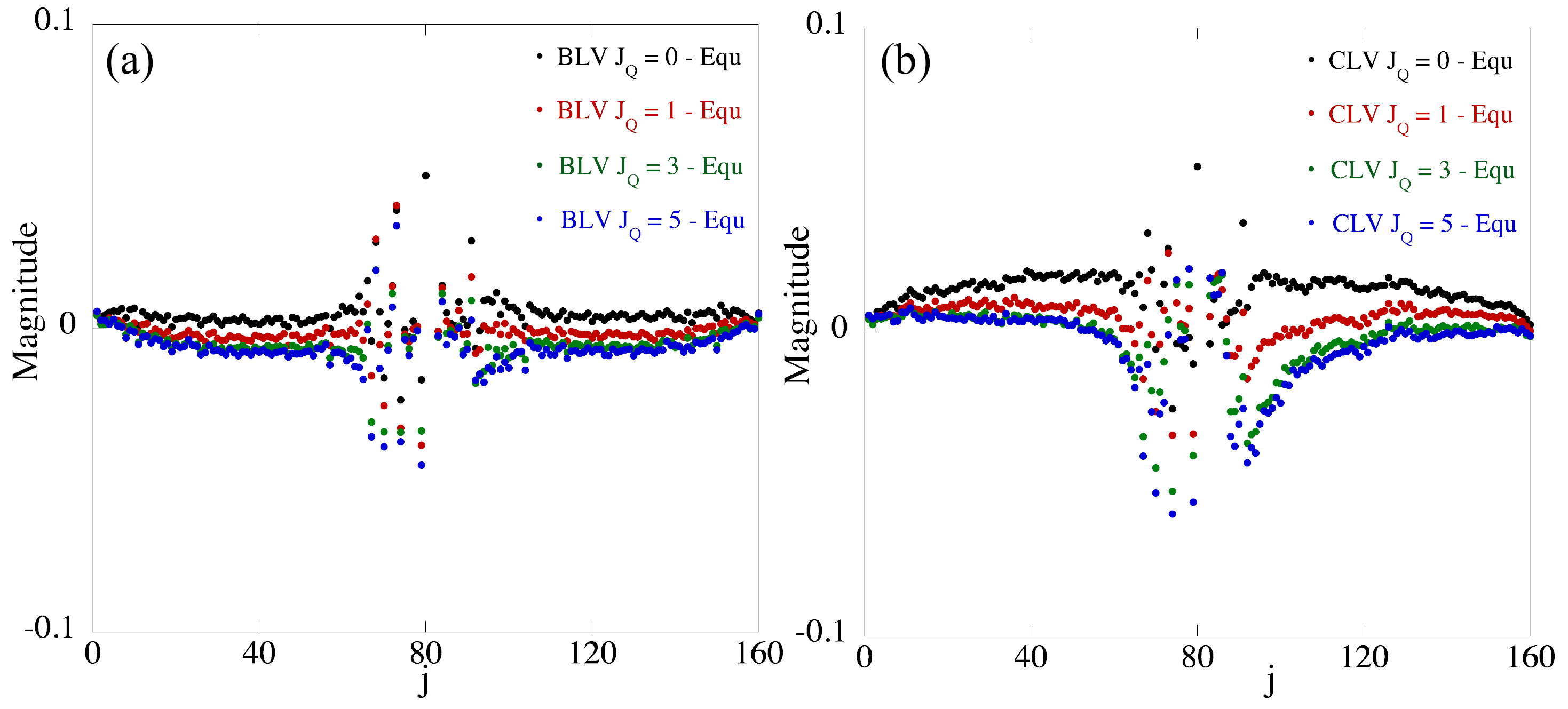}
                \caption{The difference in the average BLV and CLV localisations between equilibrium and nonequilibrium systems $\big< L^{(j)} \big>_{J_Q} - \big< L^{(j)} \big>_{Eq}$ for $\rho = 0.8$ and $N=40$, panel (a) for the BLVs and panel (b) for the CLVs. Both figures are shown over the same range.}
\label{fig:locdiffshigh}
\end{figure*}

   Fig. \ref{fig:locdiffshigh} shows the difference in the average localisation for equilibrium and nonequilibrium $\rho=0.8$ systems. Fig. \ref{fig:locdiffshigh} (a) shows the difference for the BLVs 
   while Fig. \ref{fig:locdiffshigh} (b) shows the difference for the CLVs. 
   
   There are relatively small differences between the positive and negative continuous and highly localised BLVs and CLVs, the most significant differences are in the Lyapunov mode region. For the $J_Q=0$ system the localisation measure maintains a small increase across most of the vectors. This steadily decreases with increasing heat current but very little difference is seen for the $J_Q=3$ and $J_Q=5$ systems (for both the BLVs and CLVs).
   The CLVs change more than the BLVs as the heat current increases.
   
   The differences for vectors in the central mode region are dependent on the mode characteristics. Beyond the scale of the figure are two outlier points belonging to the negative zero modes. The large changes are due to the $\gZ^{-1}$ and $\gZ^{-2}$ modes swapping their position; their localisation measure contributions also swap and leads to a large calculated difference. For the $\vZ^{-1}$ and $\vZ^{-2}$ modes, both are seen to increase their localisation measure values, due to the small contributions left in the $\delta p_x$ and $\delta p_y$ components of the modes (as discussed above).
   The largest differences to the localisation measure occur within the mode region for the positive Lyapunov modes that undergo mode mixing (both BLV and CLV), with the modes becoming more localised. 
      
\begin{figure*}
                \includegraphics[width=\textwidth]{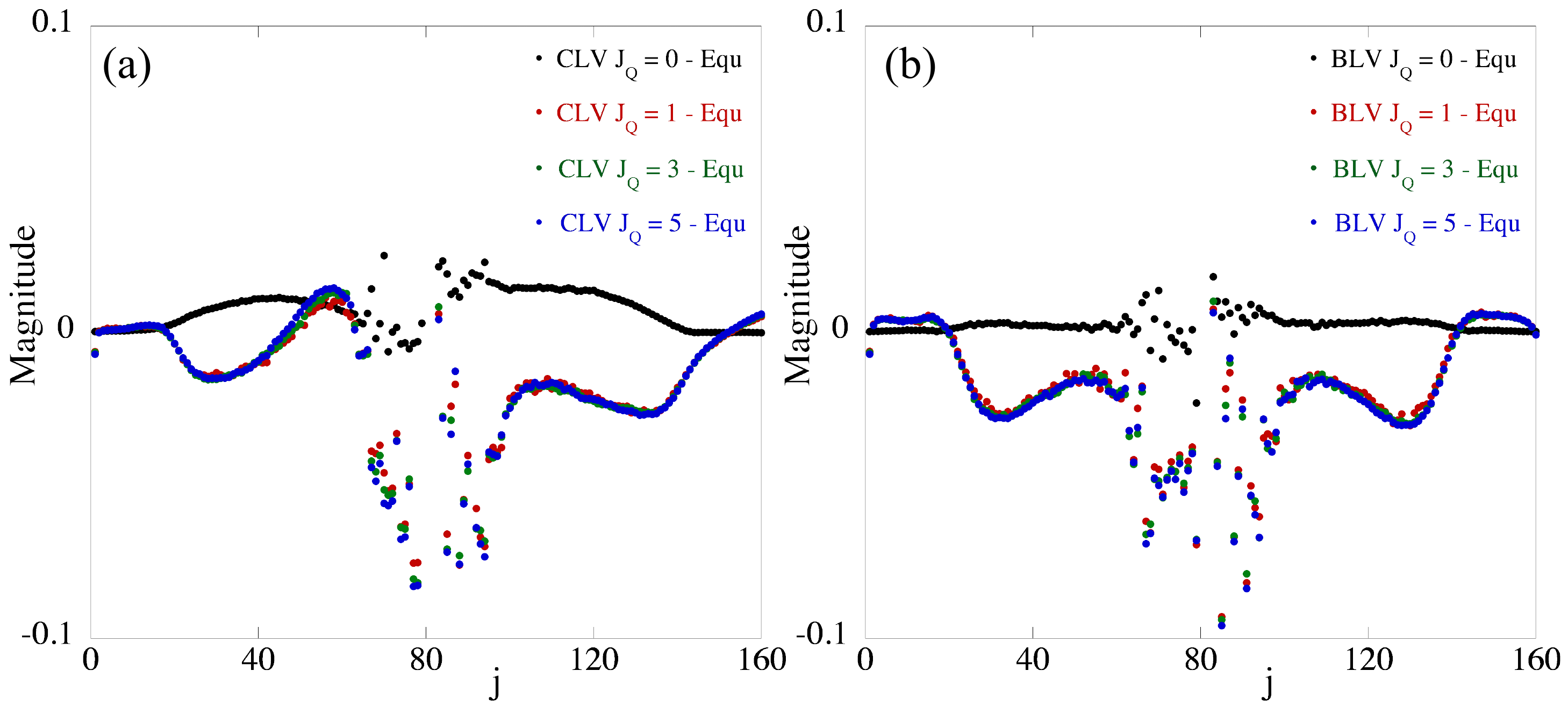}
                \caption[Low Density Localisation Differences]{The difference in the average BLV and CLV localisations between equilibrium and nonequilibrium systems $\big< L^{(j)} \big>_{J_Q} - \big< L^{(j)} \big>_{Eq}$ for $\rho = 0.003$ and $N=40$, panel (a) for the BLVs and panel (b) for the CLVs. Again, both panels are shown over the same range.}
\label{fig:locdiffslow}
\end{figure*}

   The localisation measures of the equilibrium vectors decrease significantly with decreasing density. Fig. \ref{fig:locdiffslow} shows the difference in the average localisation for the $\rho=0.003$ systems with increasing heat current over the same range as for the high density systems. Fig. \ref{fig:locdiffslow} (a) shows the differences for the BLVs 
   while panel \ref{fig:locdiffslow} (b) shows the difference for the CLVs. 

   Far clearer differences are seen here, with a clear distinction between the $J_Q=0$ and $J_Q \neq 0$ differences to the equilibrium localisation. For the $J_Q \neq 0$ systems, the differences in localisation are essentially independent of the magnitude of the applied heat current, both for the BLVs and CLVs.
   
   For the BLVs, the most localised positive and negative vectors are seen to remain invariant in the $J_Q=0$ system and show small increases in the $J_Q \neq 0$ systems. The positive and negative continuous region vectors also remain invariant in the $J_Q=0$ system, but show clear decreases in the $J_Q \neq 0$ systems.

   Unlike the symmetric BLV localisation differences, the conjugate CLVs do not give identical differences. The most localised positive CLVs remain invariant regardless of the heat current, due to the asymptotic strong localisation \cite{Morr2012}. The positive and negative continuous region CLVs become less localised for the $J_Q=0$ system. For the $J_Q \neq 0$ systems, the positive continuous region CLVs localisation first decreases then steadily increase towards the mode region, while the negative continuous region CLV maintain the decrease of their localisation towards the mode region. 

\begin{figure*}
                \includegraphics[width=\textwidth]{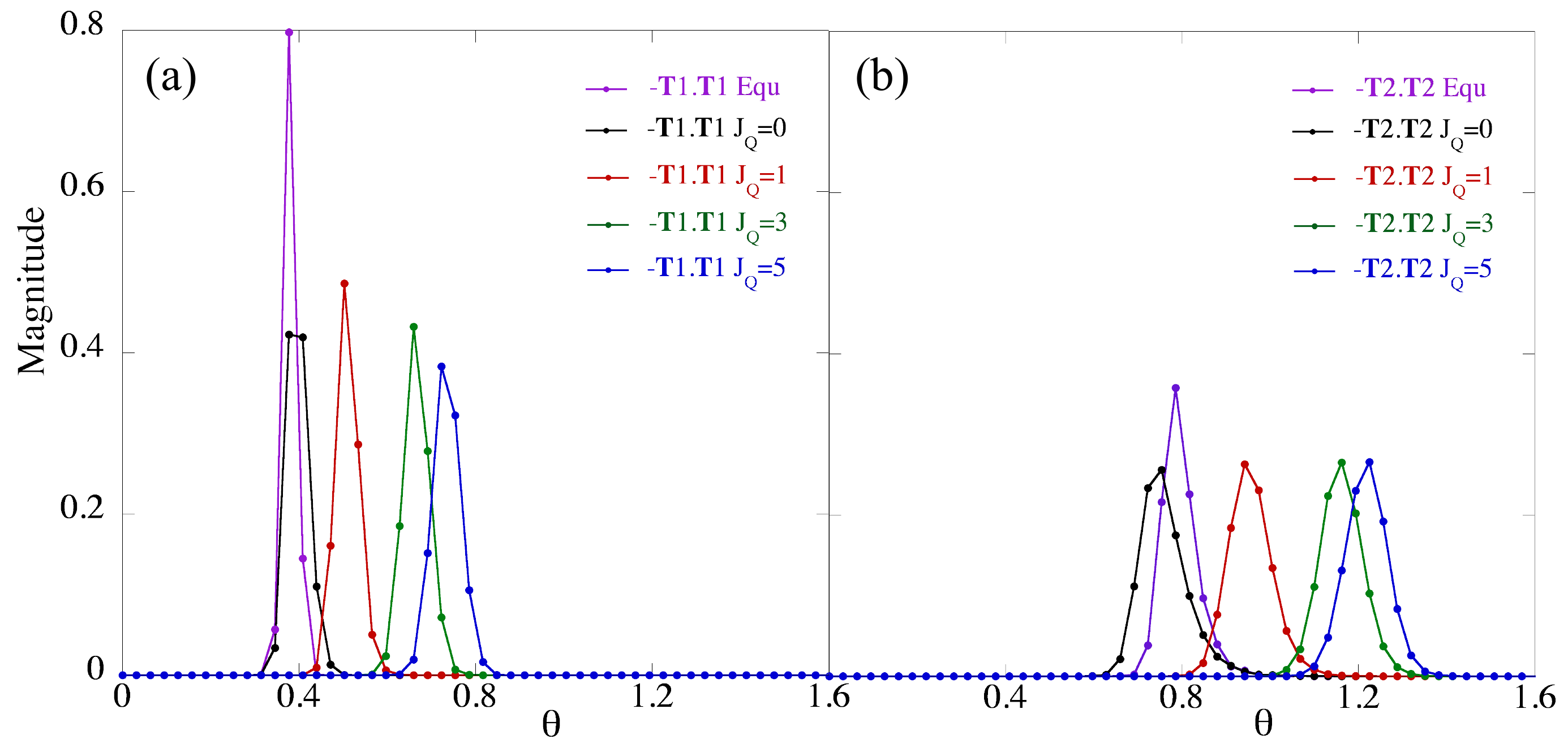}
                \caption{The angles distributions between the $\vT^1$ and $\vT^2$ conjugate pairs for equilibrium and nonequilibrium $N=40$ and $\rho=0.8$ systems, panel (a) for the $\vT^1$ modes and panel (b) for the $\vT^2$ modes. The increase in the peak angle towards $\pi/2$ is due to the increase in the $\lambda^{T}$ of the modes.}
\label{fig:heattmodes}
\end{figure*}
      
   The central mode region shows the largest and most varied differences, like for the $\rho=0.8$ systems. The swapping of the $\gZ^{-1}$ and $\gZ^{-2}$ modes produce the same large calculated difference, also seen for the $\vZ^{-1}$ and $\vZ^{-2}$ modes. The localisation measure of the positive and negative BLV Lyapunov modes remains essentially invariant for the $J_Q=0$ system.  For the $J_Q \neq 0$ systems, the positive modes decrease almost independent of the applied heat current, while the negative modes fluctuate dependent on their mode characteristics. 
   The localisation measure of the positive CLV Lyapunov modes mirrors that of the BLV modes for the $J_Q=0$ system, while the negative CLV modes become more delocalised as they take both coordinate and momentum contributions in their functional forms. 

   The most important aspect of these figures is, unlike the equilibrium BLV and CLV localisation and the majority of the nonequilibrium BLV localisations, the conjugate nonequilibrium CLV localisations are no longer symmetric. Time reversal symmetry at equilibrium assured that although made of different forms, the conjugate CLVs had identical localisation (see \cite{Morr2012}). Here we see that in nonequilibrium systems this symmetry is no longer present; Figs. \ref{fig:locdiffshigh} (b) and \ref{fig:locdiffslow} (b) show that as well as the conjugate CLVs containing different forms, they also show different average localisation properties 
   
\subsection{Angle Separations}\label{subsec:NEAS}   
   
   A crucial question to ask is: does the breaking of energy conservation alter the stable and unstable manifolds of the tangent space? It is known that the separation between the conjugate hydrodynamic modes approaches $\pi/2$ with increasing $n$ and maintains orthogonality once reached (see \cite{TruMor2011}). The $N=100$ and $N=40$ equilibrium systems have equivalent angle separations between conjugate pairs (for modes with the same exponent value), therefore they will give the same peak angle. With the introduction of a heat current the magnitude of all exponents increases, which in turn moves their peak angles towards $\pi/2$, and can force the angle between conjugate pairs into orthogonality sooner in comparison to their previously non-orthogonal separation.

\begin{table}[ht]
\caption[Conjugate Transverse Angle Predictions]{A comparison between the predicted and numerical angles between the conjugate transverse modes for equilibrium and nonequilibrium systems. The predicted angles are found using Eq. (\ref{equ:limittheta}), while the numerical angles are found from Fig. \ref{fig:heattmodes}. For systems with $N=40$ and $\rho=0.8$.}
\renewcommand{\arraystretch}{1.5}
\begin{tabular}{ccccccc}
\noalign{\hrule height 2pt}
\multirow{2}{*}{System} & \multirow{2}{*}{$\lambda^T_1$} & \multirow{2}{*}{$\tau$} & \multicolumn{2}{c}{$\cos \theta_{(-T1,T1)}$} & \multicolumn{2}{c}{$\cos \theta_{(-T2,T2)}$} \\
\cline{4-7}
  & & & Pred. & Num. & Pred. & Num. \\
\hline
Equil	&	0.1994	&	0.006485	&	0.9208	&	0.9297	&	0.6850	&	0.6375	\\
$J_Q=0$	&	0.1835	&	0.007083	&	0.9329	&	0.9245	&	0.7331	&	0.7306	\\
$J_Q=1$	&	0.2645	&	0.004851	&	0.8607	&	0.8766	&	0.4458	&	0.5897	\\
$J_Q=3$	&	0.3378	&	0.003733	&	0.7728	&	0.7769	&	0.0961	&	0.3887	\\
$J_Q=5$	&	0.3797	&	0.003292	&	0.7129	&	0.7374	&	-0.1422	&	0.3530	\\
\noalign{\hrule height 2pt}\end{tabular}
\label{tab:heattmodes}
\end{table}      

   Fig. \ref{fig:heattmodes} shows the angle between the conjugate covariant transverse modes for both the equilibrium and nonequilibrium systems, Fig. \ref{fig:heattmodes} (a) for the $\vT^1$ modes, Fig. \ref{fig:heattmodes} (b) for the $\vT^2$ modes. As the heat current increases, the peak angle of the angle separation increases towards $\pi/2$ for both the $\vT^1$ and $\vT^2$ modes. This is due to the heat current increasing the magnitude of the mode exponents uniformly (as seen in Eq. (\ref{equ:heatexpsratio})). As the $J_Q=0$ system decreases the magnitude of the exponents slightly, the peak angle is moved \emph{away} from $\pi/2$ slightly. This movement of peak angle can be predicted using an approximation first developed in \cite{TruMor2011} 
\begin{eqnarray}\label{equ:limittheta}
\cos \theta_{(-Tn,Tn)}(t \rightarrow \infty) &=& 1 - 2 {\lambda^{T}_1}^2 n^2 + 8 {\lambda^{T}_1}^3 n^3 \tau + ...
\end{eqnarray}   
   by knowing the average free flight time $\tau$ for each of the systems as well as the value of the first transverse exponent $\lambda^T_1$. 

\begin{figure*}
                \includegraphics[width=\textwidth]{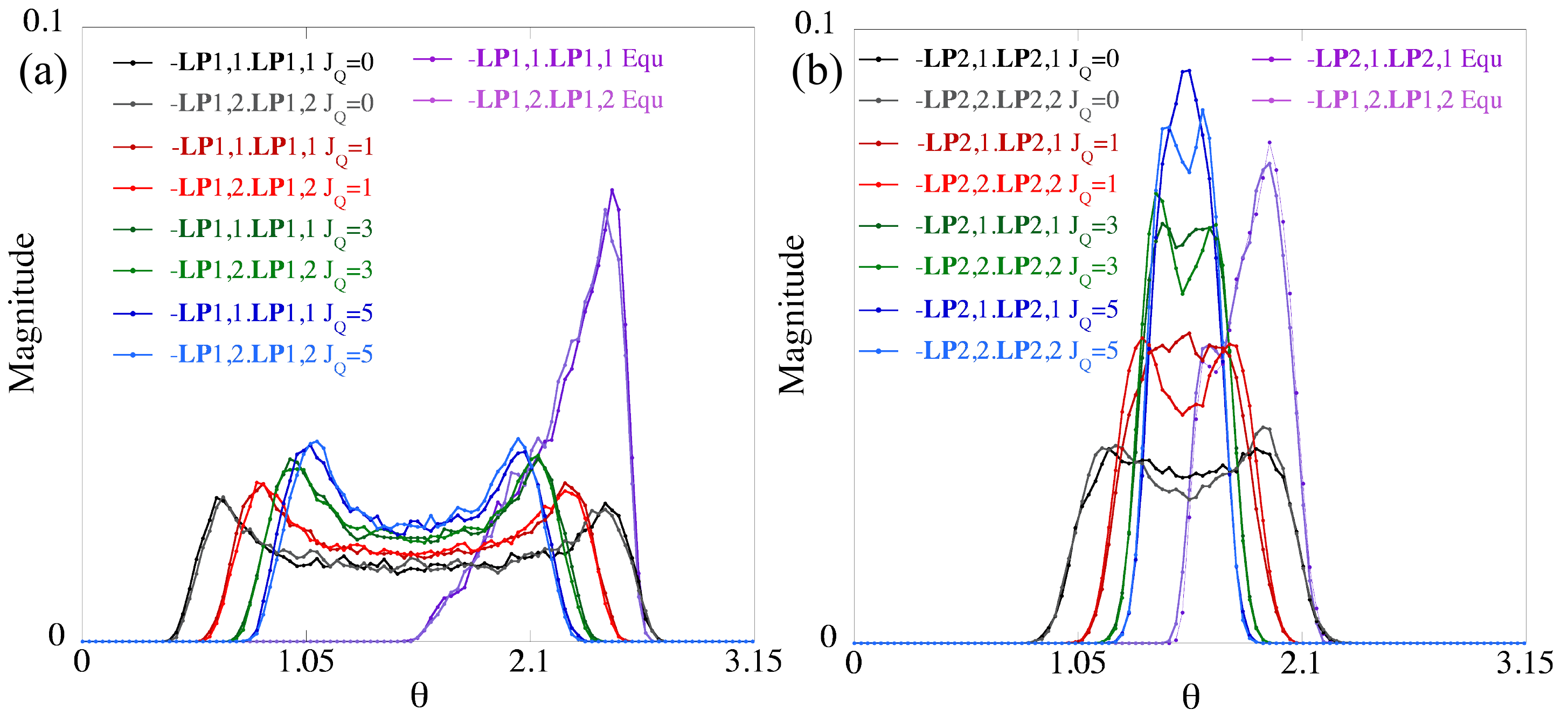}
                \caption{The angles distributions between the $\vLP^{1,i}.\vLP^{-1,i}$ and $\vLP^{2,i}.\vLP^{-2,i}$ conjugate pairs for equilibrium and nonequilibrium $N=40$ and $\rho=0.8$ systems, corresponding to the diagonal elements of the $C$ matrix, panel (a) for the $\vLP^1$ modes and panel (b) for the $\vLP^2$ modes. The modes move towards $\pi/2$ with increasing heat current.}
\label{fig:heatlpmodes}
\end{figure*}
            
   Table \ref{tab:heattmodes} shows a comparison between the predicted peak angles and the numerical peak angles for both equilibrium and nonequilibrium transverse conjugate pairs. The predicted angle is extremely accurate for the $\vT^1$ modes (for both equilibrium and nonequilibrium), within the tolerance of the histogram width used to calculate the distributions ($\sim 0.03$). The accuracy for the $\vT^2$ modes drops off with increasing heat current, with the $J_Q=5$ predicted angle implying orthogonality which is not seen numerically. The numerical angles approach $\pi/2$ much slower than the predicted angles, showing that the approximations used to find Eq. (\ref{equ:limittheta}) is inaccurate for large $\lambda^T_1$ and $n$ values, which is to be expected. 
            
   The transverse mode exponents are only scaled by the heat current and from Fig. \ref{fig:heattmodesCV} their functional forms are not altered significantly, meaning their angle distributions are not significantly changed by the breaking of energy conservation. This is not the case for the $\vLP$ modes; while the time-independent functional forms of the nonequilibrium $\gLP$ and $\vLP$ modes are not substantially altered (similar to the transverse modes), their time dependence is. 
   The angle distributions between the conjugate $\vLP$ modes are strongly bounded away from $\pi/2$ \cite{TruMor2011}. The breaking of energy conservation causes the conjugate $\vLP$ mode angle distributions to become symmetric about $\pi/2$, as seen in Fig. \ref{fig:heatlpmodes}; Fig. \ref{fig:heatlpmodes} (a) for the $\vLP^1$ modes and Fig. \ref{fig:heatlpmodes} (b) for the $\vLP^2$ modes. 
   With the introduction of a heat current the angle separation is no longer bounded by $\pi/2$, but spreads and becomes symmetric about $\pi/2$. In the same way as for the transverse modes the peak angle maintains the movement towards $\pi/2$ with increasing heat current. Symmetric peaks about $\pi/2$ can be seen for the $\vLP^1$ modes, while for the $\vLP^2$ modes the effect of the movement towards $\pi/2$ becomes more significant and the modes are forced into orthogonality. 
      
   At equilibrium the $C$ matrix element giving the angle between the conjugate $\vLP$ modes, the $f_{LP}/\sqrt{f^2_{LP}+1}$ function, starts at zero (giving an angle of $\pi/2$), moves away quickly and does not return, with the angle distribution mimicking the histogram of the evolution function values. 
   Numerically we find once energy is no longer conserved with the introduction of a heat current in nonequilibrium systems the top right of the $C$ matrix for the $\vLP$ modes 
   is altered. At equilibrium the relevant $C$ matrix elements for the $\vLP$ modes (Eq. (\ref{equ:cforLPs})) is given as 
\begin{eqnarray*}
C_{LP} = \left( \begin{array}{cccc} \cline{3-4} 1 & 0 & \multicolumn{1}{|c}{0} & \multicolumn{1}{c|}{\frac{f_{LP1}}{\sqrt{f^2_{LP1} + 1}}} \\ 0 & 1 & \multicolumn{1}{|c}{\frac{f_{LP1}}{\sqrt{f^2_{LP1} + 1}}} & \multicolumn{1}{c|}{0} \\ \cline{3-4} 0 & 0 & \frac{1}{\sqrt{f^2_{LP1} + 1}} & 0 \\ 0 &  0 &  0 & \frac{1}{\sqrt{f^2_{LP1} + 1}}\end{array}\right) \begin{array}{c} \multirow{2}{*}{$\leftarrow C^{\ast}_{LP}$} \\ \\ \\ \\ \end{array}.
\end{eqnarray*}   
   The entries in the top right of the $C_{LP}$ matrix, which we indicate with the box and label as $C^{\ast}_{LP}$, are seen to change. The $f_{LP}/\sqrt{f^2_{LP}+1}$ matrix elements (indicating the direct conjugate pair), and the previously zero value matrix elements (indicating the previously orthogonal indirect pairs within the conjugate modes), oscillate with respect to time $\pi/2$ out of phase in nonequilibrium systems as
\begin{eqnarray}
C^{\ast}_{LP} &=& \left[ \begin{array}{cc} 0 & \frac{f_{LP1}}{\sqrt{f^2_{LP1} + 1}} \\ \frac{f_{LP1}}{\sqrt{f^2_{LP1} + 1}} & 0 \end{array} \right] \nonumber \\
\label{equ:Cmatrixrotation} &\rightarrow& \left[ \begin{array}{cc} \sin{\omega_{J_Q}t} \frac{f_{LP1}}{\sqrt{f^2_{LP1} + 1}} & \cos{\omega_{J_Q}t} \frac{f_{LP1}}{\sqrt{f^2_{LP1} + 1}} \\ \cos{\omega_{J_Q}t} \frac{f_{LP1}}{\sqrt{f^2_{LP1} + 1}} & -\sin{\omega_{J_Q}t} \frac{f_{LP1}}{\sqrt{f^2_{LP1} + 1}} \end{array} \right], 
\end{eqnarray}
   giving an oscillation within the positive $\vLP$ mode subspace (these $C^{\ast}_{LP}$ entries correspond to the $C$ matrix entries outlined in the top right $2 \times 2$ blue square of Fig. \ref{fig:heatcmatrix}). This poses the question, where does this oscillation come from and what is the relation between $\omega_{J_Q}$ and $\omega_{n}$ (the $\vLP$ and $\gLP$ mode frequency)? 

\begin{table*}[ht]
\caption[$\LP$ Mode and $C^{\ast}_{LP}$ Matrix Frequencies]{A comparison between the period of oscillations between the equilibrium and nonequilibrium positive and negative $\LP$ modes ($T_1$ and $T_{-1}$) and the oscillation of the heat current $C^{\ast}_{LP}$ matrix elements $T_{J_Q}$ in terms of collision numbers of each system. The frequency of oscillations for these periods is found via $\omega_i = 2\pi/T_i\tau$ with $\tau$ given in Table \ref{tab:heattmodes}. The frequency difference between the positive and negative modes $\omega_{1} - \omega_{-1}$ gives the frequency of the $C^{\ast}_{LP}$ matrix oscillations $\omega_{J_Q}$. For systems with $N=40$ and $\rho=0.8$.}
\renewcommand{\arraystretch}{1.5}
\begin{tabular}{cccccccccc}
\noalign{\hrule height 2pt}
\multirow{2}{*}{System} & \multicolumn{2}{c}{$T_{1}$ of $\LP^{1}$} & \multicolumn{2}{c}{$T_{-1}$ of $\LP^{-1}$} & $T_{J_Q}$ & \multirow{2}{*}{$\omega_{1} - \omega_{-1}$} & \multirow{2}{*}{$\omega_{J_Q}$} \\
\cline{2-5}
  &  BLV  & CLV  & BLV  & CLV  & ($\pm$ 800) & & \\
\hline
Equil	& 2900 & 2900 & 2900 & 2900 & - & - & - \\
$J_Q=0$ & 2600 & 2600 & 3000 & 3000 & 18800 & 0.0455 & 0.047 $\pm$ 0.002 \\
$J_Q=1$	& 2700 & 2700 & 3150 & 3150 & 17100 & 0.0685 & 0.075 $\pm$ 0.004 \\
$J_Q=3$	& 2700 & 2700 & 3200 & 3200 & 17800 & 0.0974 & 0.095 $\pm$ 0.004 \\
$J_Q=5$	& 2700 & 2700 & 3250 & 3250 & 16700 & 0.1196 & 0.114 $\pm$ 0.006 \\
\noalign{\hrule height 2pt}\end{tabular}
\label{tab:heatfreq}
\end{table*}
   
   Table \ref{tab:heatfreq} shows a comparison between the period of oscillation between the equilibrium and nonequilibrium $\LP$ modes for both positive and negative  ($T_1$ and $T_{-1}$) BLV and CLV modes, and the period of oscillation of the nonequilibrium $C^{\ast}_{LP}$ matrix elements $T_{J_Q}$ in terms of collision numbers of each system. The frequencies of these oscillations can be found from the period as $\omega_i = 2\pi/T_i\tau$, where $\tau$ differs for each system due to the applied heat current (see Table \ref{tab:heattmodes}). Using the period of oscillation in terms of collision numbers is more general and removes the system dependent time factor. From Table \ref{tab:heatfreq} we see that for nonequilibrium systems the period of oscillation for the negative $\LP$ modes (both BLV and CLV) \emph{increases} in comparison to the equilibrium period, while the period of oscillation for the positive $\LP$ modes (both BLV and CLV) \emph{decreases} in comparison to the equilibrium period, a feature previously witnessed \cite{TanMor2007a}. The change in period of the positive modes appears essentially independent of the applied heat current, while there is a small increase with heat current for the negative modes. The $C^{\ast}_{LP}$ matrix oscillation period is difficult to determine as there are substantial fluctuations in the values over time, hence the large error. 

\begin{figure*}[htbp]
   \includegraphics[width=0.85\textwidth]{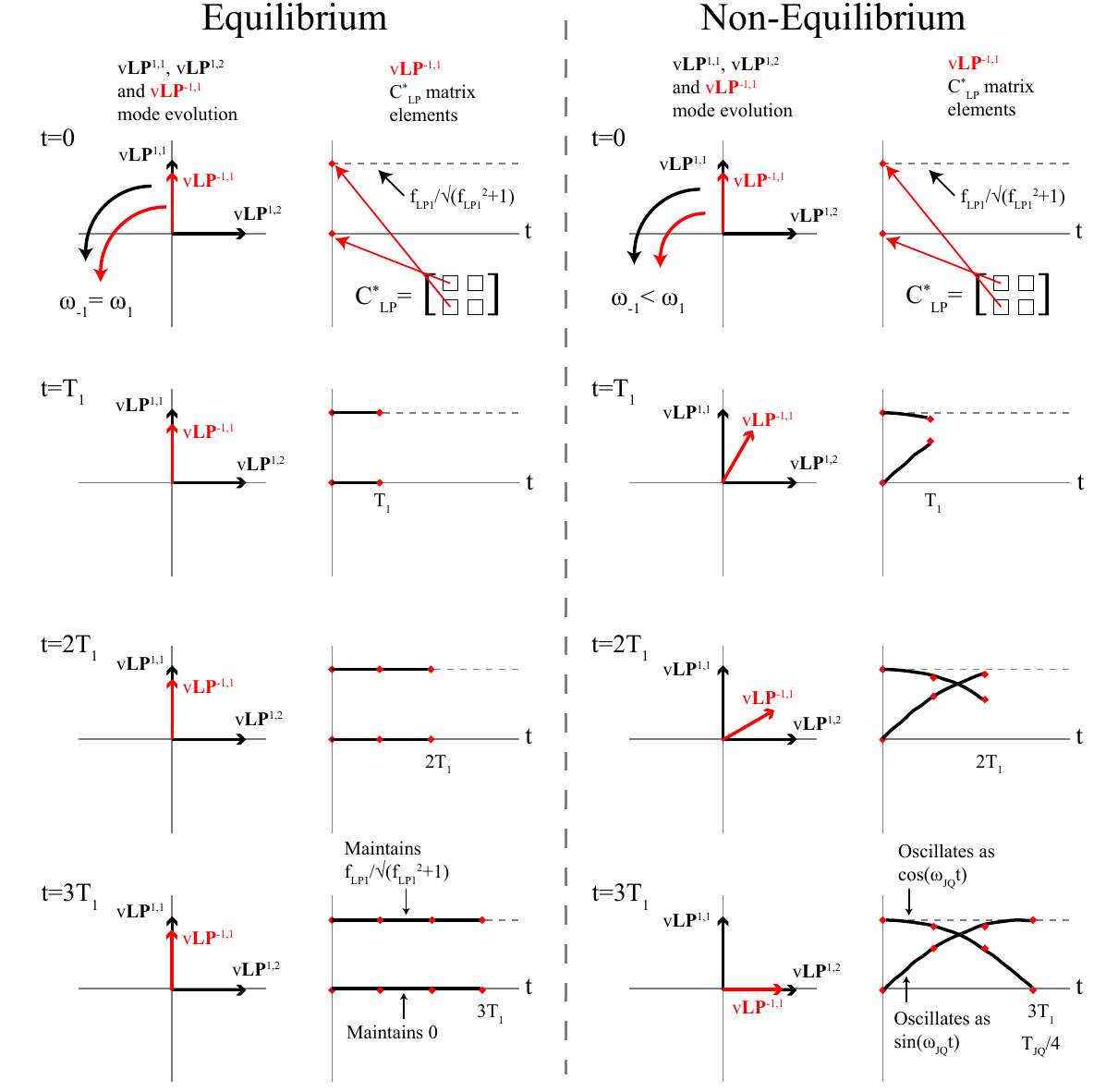}
\caption[Covariant $\LP$ Mode Oscillation Explanation]{In equilibrium systems the positive and negative $\LP$ mode frequencies are equal, therefore the $C^{\ast}_{LP}$ matrix entries of the $\vLP^{-1,1}$ mode maintain steady values (the plotted entries corresponding to the left hand side of Eq. (\ref{equ:Cmatrixrotation})). In nonequilibrium systems, the negative $\vLP^{-1,1}$ (and therefore $\gLP^{-1,1}$) mode oscillates slower than the positive modes $\omega_{-1} < \omega_{1}$ (from Table \ref{tab:heatfreq}), therefore the negative mode oscillates within the subspace of the positive $\vLP^{1,1} \vLP^{1,2}$ mode pair at a rate determined by the difference in the positive and negative frequencies $\omega_{J_Q} = \omega_{1} - \omega_{-1}$. This causes the $C^{\ast}_{LP}$ matrix entries of the $\vLP^{-1,1}$ mode to oscillate at this $\omega_{J_Q}$ frequency (the plotted entries corresponding to the right hand side of Eq. (\ref{equ:Cmatrixrotation})). The periods here are indicative only and serve as illustrative approximations.}
\label{fig:LPmodeosc}
\end{figure*}

   By finding the frequency of the $\vLP^1$ and $\vLP^{-1}$ mode oscillations (which are equal to the $\gLP^1$ and $\gLP^{-1}$ frequencies) for each system we can see how the $C^{\ast}_{LP}$ matrix oscillation forms. As the positive and negative nonequilibrium modes oscillate at different frequencies, over time this will lead to a separation between the direct conjugate modes. Taking the difference between the positive and negative frequencies $\omega_{1} - \omega_{-1}$ gives this separation frequency (a beat frequency, shown in Table \ref{tab:heatfreq}). At some times each negative vector in the $\vLP$ mode pair will be pointing in the direction of its direct conjugate vector (for example, $\vLP^{1,1}$ for $\vLP^{-1,1}$) but due to the frequency difference at other times it will be pointing in the direction of the previously time-orthogonal degenerate pair vector (for example, $\vLP^{1,2}$ for $\vLP^{-1,1}$), shown visually in Fig. \ref{fig:LPmodeosc}. The rate at which this oscillation occurs is $\omega_{1} - \omega_{-1}$. 
   In the last column of Table \ref{tab:heatfreq} we see that this separating frequency difference is what accounts for the $C^{\ast}_{LP}$ matrix oscillation $\omega_{1} - \omega_{-1} = \omega_{J_Q}$. 
    
   As shown in Fig. \ref{fig:LPmodeosc}, in nonequilibrium systems the difference between the positive and negative $\vLP$ mode frequencies causes each negative vector to slowly move between having the functional forms of each vector in the positive degenerate $\vLP^{1}$ mode pair (which are $\pi/2$ out of phase in time, see Section \ref{equilanal}). This means the $C^{\ast}_{LP}$ matrix elements relating to the two vectors in the negative $\vLP^{-1}$ mode (which indicates the functional form each negative vector is comprised of) oscillates about zero (as $\sin{\omega_{J_Q}t} $ or $\cos{\omega_{J_Q}t} $ from Eq. (\ref{equ:Cmatrixrotation})), the rate at which this oscillation occurs given by the positive and negative frequency difference $\omega_{1} - \omega_{-1} = \omega_{J_Q}$. This $C^{\ast}_{LP}$ matrix oscillation in turn forces the angle distribution between the modes (formed from the distribution of the $C$ matrix elements) to now oscillate about $\pi/2$ and forms the symmetric distributions seen in Figs. \ref{fig:heatlpmodes} (a) and \ref{fig:heatlpmodes} (b). 

   t
      
   This $C^{\ast}_{LP}$ matrix oscillation would also change the angle distributions between the indirect $\vLP^{-1,1}$ and $\vLP^{1,2}$ pair (previously orthogonal from the equilibrium $C^{\ast}_{LP}$ matrix elements in Eq. (\ref{equ:Cmatrixrotation})) into an equivalent distribution as given in Fig. \ref{fig:heatlpmodes} (a). The $C^{\ast}_{LP}$ matrix oscillation (and hence symmetric angle distribution) is not seen at equilibrium because the positive and negative $\vLP$ mode frequencies are equal.
   
   Fig. \ref{fig:heatlpmodes} (b) shows the increase in the magnitude of the exponent forces the $\vLP^2$ mode - which was bounded away from $\pi/2$ in the equilibrium case - into orthogonality, independent of the oscillation of the $C^{\ast}_{LP}$ matrix elements. In this way the $\vLP^2$ modes mimics the angle distributions of the continuous region vectors already witnessed \cite{TruMor2011,Morr2012}. The position of the mode within the spectrum does not change with the increase in heat current, therefore the position of the peak angle depends only on the magnitude of the exponent.
   
   Combining this movement in peak angle with the $C^{\ast}_{LP}$ matrix oscillation due to the $\vLP$ mode frequency difference describes in full the numerically observed angles distributions between the conjugate $\vLP$ modes, their symmetry about $\pi/2$ for nonequilibrium systems and their movement towards orthogonality with increasing heat current.
   
   From this we can conclude with the increasing magnitude of the exponents, more of the conjugate manifolds are forced into orthogonalisation with each other. The only stable and unstable manifolds that can form tangencies therefore are the smallest magnitude conjugate pairs, the number of which become vanishingly small as the heat current increases.

\section{Conclusion}
   
   We have shown that the changes of the Lyapunov exponents and Lyapunov modes with the introduction of a heat current can be explained. 
   Eqs. (\ref{equ:heatexpsratio}) to (\ref{equ:heatexpsratio3}) show that the Lyapunov exponents of any nonequilibrium system are found from a scaling of the equilibrium exponents as a power law in heat current. 
   The change in the functional form of the nonequilibrium backwards zero modes are a result of the requirement they remain orthogonal to each other, a requirement not present for the nonequilibrium covariant zero modes. 
   Table \ref{tab:heattmodes} shows that, from the knowledge of the nonequilibrium exponents (found from the equilibrium exponents), the converged angle between the nonequilibrium hydrodynamic covariant conjugate Lyapunov modes can be predicted. 
   The asymmetry between the positive and negative nonequilibrium $\LP$ mode frequencies lead to the negative mode oscillating between the functional forms of each orthogonal mode in the positive conjugate mode pair. 
   This in turn causes the angle distributions between the conjugate $\vLP$ modes to oscillate symmetrically about $\pi/2$ at a rate given by the difference between the positive and negative mode frequencies.


\bibliography{NonEquBib.bib}


\end{document}